\documentclass[a4paper, oneside, english, 11pt]{report}

\usepackage[letterpaper,top=2cm,bottom=2cm,left=3cm,right=3cm,marginparwidth=1.75cm]{geometry}
\usepackage[final]{pdfpages}
\usepackage{amsmath,amsthm}
\usepackage{amsfonts} 
\usepackage{graphicx}
\usepackage[colorlinks=true, allcolors=blue]{hyperref}
\usepackage{comment}
\usepackage[T1]{fontenc}
\usepackage[utf8]{inputenc}
\usepackage{booktabs}
\usepackage{hyperref}
\usepackage[sorting=none]{biblatex}
\usepackage{wrapfig}
\usepackage{marginnote}

\usepackage{amssymb}
\usepackage{afterpage}
\usepackage{array}

\theoremstyle{definition}

\newcommand{\iprod}{\mathbin{\lrcorner}}
\newcommand{\ds}{\displaystyle}

\newcommand{\paren}[1]{{\ensuremath{\left ( #1 \right )}}}

\newcommand{\WXL}{\ensuremath{W_{XL}}}
\newcommand{\WXX}{\ensuremath{W_{XX}}}
\newcommand{\WLX}{\ensuremath{W_{LX}}}
\newcommand{\WLL}{\ensuremath{W_{LL}}}
\newcommand{\UD}[2]{\ensuremath{^{#1}_{\phantom{#1} #2}}}

\newcommand{\set}[1]{\left \{ #1 \right \} }

\newcommand{\R}{\mathbb R}

\newcommand{\tuple}[1]{\ensuremath{\left (#1 \right )}}

\newcommand{\struct}[1]{\ensuremath{\left (#1 \right )}}

\newcommand{\map}[2]{\ensuremath{ #1 \left ( #2 \right )}}

\newcommand{\valat}[2]{\ensuremath{ \left. #1 \right |_{#2}}}

\newcommand{\DU}[2]{\ensuremath{_{#1}^{\phantom{#1} #2}}}

\newcommand{\UDDD}[4]{\ensuremath{^{#1}_{\phantom{#1} #2 #3 #4}}}

\newcommand{\calL}{\ensuremath{\mathcal{L}}}
\newcommand\blankpage{%
    \null
    \thispagestyle{empty}%
    \addtocounter{page}{-1}%
    \newpage}

\begin{document}

\makeatletter
\begin{titlepage}
    \begin{center}
        \hrule
        \vspace*{1cm}
        \Huge
        \textbf{Bilocal geodesic operators as a tool of investigating the optical properties of spacetimes \linebreak (updated version)}
        \vspace{1cm}
        \hrule
        \vspace{0.5cm}
        \begin{figure}[htbp]
             \centering
             \includegraphics[width=.3\linewidth]{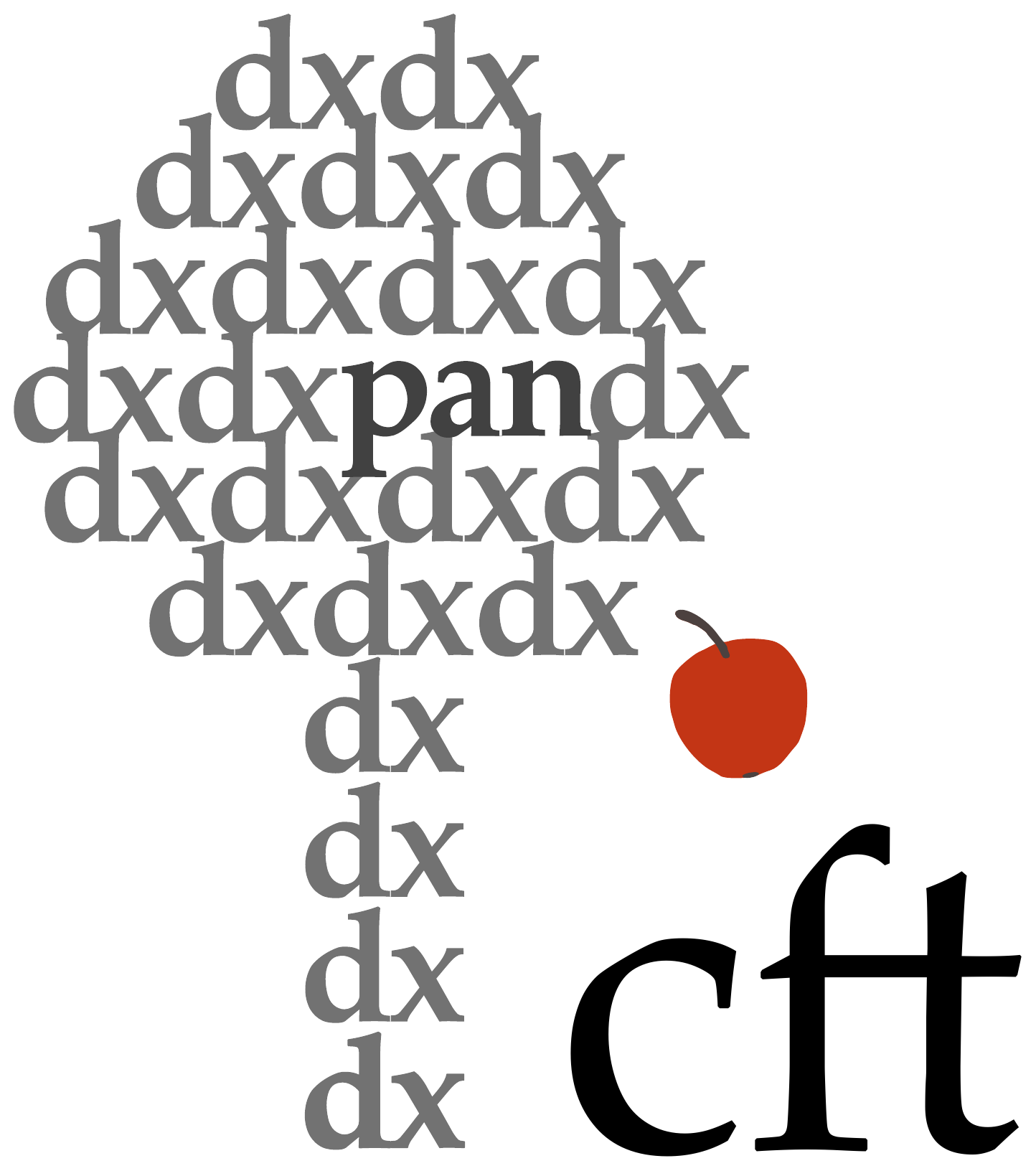}
        \end{figure}

        \vspace{1.5cm}

        \large
        \textbf{Author}: \Large{Julius Serbenta}\\
        \large
        \textbf{Supervisor}: \Large{Prof. Mikołaj Korzyński}\\

        \vspace{1cm}
        \Large{\textbf {Centrum Fizyki Teoretycznej Polskiej Akademii Nauk}}\\
        \vspace{1cm}
       \textit{A thesis submitted in partial fulfillment of the requirements for the degree of}\\
       \textit{Doctor of Philosophy in Physics}\\
       \textit{}\\
        
        \vspace{1cm}
        {\Large \today}

    \end{center}
\end{titlepage}
\makeatother

\blankpage{}
\thispagestyle{empty}%
\topskip0pt
\vspace*{\fill}
\begin{center}
    \textit{Dedicated to my grandmother}
\end{center}
\vspace*{\fill}
\blankpage{}
\pagenumbering{roman}
\topskip0pt
\vspace{-1cm}
\begin{center}
    \chapter*{Abstract}
\end{center}
\addcontentsline{toc}{chapter}{Abstract}

	In my thesis, I present one particular example of the formalism capable of describing the propagation of a family of light rays in a curved spacetime. It is based on the resolvent operator of the geodesic deviation equation for null geodesics which is known as the \textit{bilocal geodesic operator} (BGO) formalism. The BGO formalism generalizes the standard treatment of light ray bundles by allowing observations extended in time or performed by a family of neighbouring observers. Furthermore, it provides a more unified picture of relativistic geometrical optics and imposes a number of consistency requirements between the optical observables.
	
	The thesis begins with a brief introduction of the transfer matrix and its relativistic versions known as the Jacobi propagators and the bilocal geodesic operators. A brief literature review is given illustrating various interpretations of bilocal operators in contexts of extended objects, gravitational waves, and seismology.
	
	The second chapter is dedicated to the basics of differential geometry with an emphasis on the geometry of the tangent bundle, which will later provide a foundation for the BGO formalism. We start from the coordinate systems on the base manifold and its tangent bundle and then study coordinate-dependent and independent representations of induced higher-dimensional vectors. Next, we discuss the notion of the geodesic flow and define the BGOs in terms of this flow. Finally, we display how certain results obtained on the tangent bundle lead to the differential equations for BGOs.
	
	In the third chapter, I present my original work on a fully analytical derivation of the BGOs for static spherically-symmetric spacetimes. Firstly, I summarize two different techniques to obtain an exact solution, both resting upon symmetries of the spacetime and integrability of geodesic (deviation) equations.  The methods are then applied to derive the solution both in coordinate and parallel-transported frames. Finally, the results are used to study optical distance measures in Schwarzschild spacetime.
	
	In the fourth chapter, I present several theorems about the inequality concerning optical distance measures. The result is valid irrespective of spacetime symmetries or lack thereof and depends on the validity of General Relativity together with rather standard assumptions about the matter content and propagation of light in the Universe. The chapter concludes with a short discussion about the possibility of experimental verification or rejection of the mathematical result.
	
	In the last chapter, I summarize the content of the thesis and ponder its possible extensions.	
\clearpage
\vspace{-1cm}
\begin{center}
    \chapter*{Streszczenie}
\end{center}
\addcontentsline{toc}{chapter}{Streszczenie}
 W pracy prezentuję formalizm opisujący propagację wiązek promieni światła, nazywany formalizmem \emph{bi-lokalnych operatorów geodezyjnych} (BGO). Jego podstawą jest rezolwenta równania dewiacji geodezyjnych dla geodezyjnych zerowych. Formalizm BGO uogólnia standardowy opis wiązek światła, pozwalając na obserwacje rozciągające się w czasie lub wykonane przez rodzinę  obserwatorów znajdujących się blisko siebie. Ponadto wprowadza on ujednolicony opis relatywistycznej optyki geometrycznej, co pozwala udowodnić  ścisłe relacje między obserwablami i krzywizną czasoprzestrzeni.

Dysertacja rozpoczyna się krótkim wprowadzeniem do tematyki macierzy przejścia (transfer matrix) i jej relatywistycznych uogólnień, zwanych czasami propagatorami Jacobiego lub bi-lokalnymi operatorami geodezyjnymi. Podaję też krótki przegląd literatury ilustujący rozmaite zastosowania tych opertatrów w kontekście równań ruchu rozciągłych ciał w ogólnej teorii względności, fal grawitacyjnych i sejsmologii.

Drugi rozdział poświęcony jest matematycznym postawom geometrii różniczkowej, przede wszystkim geometrii wiązki stycznej. Materiał ten będzie potem podstawą formalizmu BGO. Rozpoczynam od przypomnienia pojęcia układu współrzędnych na czasoprzestrzeni i na wiązce stycznej oraz opisuję zależne i niezależne od 
układu współrzędnych metody rozkładu wektorów stycznych do wiązki stycznej. Natępnie opisuję analog kongruencji geodezyjnych na wiązce stycznej, zwany geodezyjnym przepływem (\emph{geodesic flow}) i definiuję przy jego pomocy bilokalne operatory geodezyjne. Na koniec wyprowadzam równania różniczkowe na te obiekty korzystając z geometrii wiązki stycznej.

W trzecim rozdziale prezentuję swoją oryginalną pracę, w której wyprowadzam dokładne wyrażnenia na BGO dla statycznych, sferycznie symetrycznych czasoprzestrzeni. Opisuję najpierw dwie techniki otrzymywania dokładnych rozwiązań, obie korzystające z wielkości zachowanych i całkowalności równań geodezyjnych w sferycznie symetrycznych czasoprzestrzeniach. Metody te stosuję potem do wyprowadzenia rozwiązań wyrażonych w reperze współrzędnościowym i transportowanym równolegle. Na koniec, korzystając z tych wyników, badam optyczne miary odległości (odległość paralaktyczną i rozmiaru kątowego) na czasoprzestrzeni Schwarzschilda.

W czwartym rozdziale prezentuję dwa ogólne twierdzenia dotycząnie nierówności między optycznymi miarami odległości. Są one prawdziwe bez względu na to, czy czasoprzestrzeń jest symetryczna, czy nie. Zakładają one prawdziwość 
ogólnej teorii względności,
standardowe warunki na rozkład materii oraz przybliżenie optyki geometrycznej dla propagacji światła. Na końcu rozdziału dystkutuję pokrótce możliwość
eksperymentalnej weryfikcacji tej nierówności.

W ostatnim rozdziale podsumowuję treść dysertacji i rozważam rozmaite możliwe uogólnienia wyników. 

\clearpage
\vspace{-1cm}
\begin{center}
    \chapter*{Declaration}
\end{center}
\addcontentsline{toc}{chapter}{Declaration}

The work presented in this thesis was completed between October 2017 and November 2022 while the author was a research student under the supervision of Prof. Mikołaj Korzyński at the Center for Theoretical Physics, Polish Academy of Sciences. The coursework was completed between October 2017 and July 2022 at the Institute of Physics, Polish Academy of Sciences, and Warsaw University. 
No part of this thesis has been submitted for any other degree at the Center for Theoretical Physics, Polish Academy of Sciences, or any other scientific institution.
\vspace{0.02\textwidth}

The thesis is based on the following papers:
\vspace{0.01\textwidth}

\begin{enumerate}
    \item Chapter~\ref{ssss}: \textbf{J. Serbenta} and M. Korzyński, \emph{Bilocal geodesic operators in static spherically-symmetric spacetimes}, Class. Quantum Grav., vol 39, 155002, 2022 \cite{serbenta2022},
    
    \item Chapter~\ref{testn}: M. Korzyński and \textbf{J. Serbenta}, \emph{Testing the null energy condition with precise distance measurements}, Phys. Rev. D, 105, 084017, 2022 \cite{korzynski2022}.
    
\end{enumerate}

\vspace{0.02\textwidth}

Descriptions of the contributions of the authors are given separately before each publication in the respective chapters.

\vspace{0.02\textwidth}
In addition to the work presented in this thesis, the author has also contributed to the following closely related articles:
\begin{enumerate}
    \item[3.] M. Grasso, M. Korzyński, and \textbf{J. Serbenta}, \emph{Geometric optics in general relativity using bilocal operators}, Phys. Rev. D, vol. 99, 064038, 2019 \cite{korzynski2021},
    \item[4.] M. Korzyński, J. Miśkiewicz, and \textbf{J. Serbenta}, \emph{Weighing the spacetime along the line of sight using times of arrival of electromagnetic signals}, Phys. Rev. D, vol. 104, 024026, 2021 \cite{grasso2019}.
\end{enumerate}
\clearpage

\begin{center}
\chapter*{Acknowledgements}    
\end{center}
\addcontentsline{toc}{chapter}{Acknowledgements}

During my doctoral studies, I met many people who influenced me in lots of different ways, whether they shared just an interesting fact about physics or allowed me to experience their culture and traditions, and discuss various philosophies of life and nature.

First and foremost, I am grateful to my supervisor prof. dr. hab. Mikołaj Korzyński for all the help I received during all these years. His deep insights and attention to detail enabled me to improve my critical thinking. Moreover, I cannot thank him enough for all the support he gave me during my research whenever I was seriously stuck.

I am grateful for the opportunity to meet so many curious minds from different parts of the world. My special thanks go to Michele, Ishika, Grzegorz, Suhani, Saikruba, Katja, Shubhayan, Lorenzo, Paweł, Anjitha, Feven, Owidiusz, Tae-Hun, Bestin, Michal, and many others not limited to CTP PAS.

I am indebted to my family, who have always encouraged me during the most difficult periods of my studies. Without your faith, I would not be here.

I would also like to express my gratitude to the rest of the staff at CTP PAS. I would especially like to thank the directors of CTP PAS and my supervisor for their support during the extension of my studies. The administrative staff definitely deserves a recognition for their assistance in completing the paperwork, especially during the period of changing laws and regulations concerning higher education.

Finally, I would like to acknowledge the support of NCN Grant No. 2016/22/E/ST9/00578 and the taxpayers whose financial contribution made these studies possible.

\pagebreak
\clearpage
\thispagestyle{empty}

\thispagestyle{empty}%
\pagebreak
\chapter*{Acronyms and notations}
\addcontentsline{toc}{chapter}{Acronyms and notations}

\definecolor{airforceblue}{rgb}{0.36, 0.54, 0.66}
\definecolor{bleudefrance}{rgb}{0.19, 0.55, 0.91}
\vskip -1cm

\section*{List of acronyms}
\begin{center}
\begin{tabular}{| c | c |} 
 \hline
 \textbf{Acronym} & \textbf{Meaning}  \\  
 \hline
 BGO & Bilocal Geodesic Operator  \\ 
 \hline
 GDE & Geodesic Deviation Equation  \\
 \hline
 GE & Geodesic Equation  \\
 \hline
 NEC & Null Energy Condition\\
 \hline
 ODE & Ordinary Differential Equation  \\
 \hline
 SNT & Semi-Null Tetrad  \\
 \hline
 
\end{tabular}
\end{center}

\section*{Notations and their meanings}
\hskip-0.5cm
\bgroup
\begin{tabular}{ | c | c |  } 
  \hline
  \textbf{Notation} & \textbf{Description}  \\
  \hline
  $g$ & metric tensor \\ 
  \hline
  $\nabla$ & Levi-Civita connection  \\ 
  \hline
  $\Gamma^\alpha_{~\mu\nu}$ & Christoffel symbol\\
  \hline
  $M$ & semi-Riemannian manifold  \\ 
  \hline
  $T_p M$ & tangent space at $p \in M$\\
  \hline 
  $\tuple{ \valat {\dfrac \partial {\partial x^\mu}}p}$ & coordinate basis vector of $T_p M$\\
  \hline
  $\mathbf X_p = v^\mu \valat{\dfrac{\partial}{\partial x^\mu}}p$ & element of $T_p M$\\
  \hline
  $TM$ & tangent bundle of $M$\\
  \hline
  $T_{\tuple {p, \mathbf X_p}}TM$ & tangent space to the tangent bundle at $\tuple {p, \mathbf X_p} \in TM$\\
  \hline
   $\tuple{ \valat {\dfrac \partial {\partial x^\mu}}p, \valat {\dfrac \partial {\partial v^\mu}}p}$ & induced basis vector of $T_{\tuple{p, \mathbf X_p}} TM$\\
   \hline
  $\mathbf Y_{\tuple{p, \mathbf X_p}}$ & element of $T_{\tuple {p, \mathbf X_p}} TM$\\
  \hline
  Greek indices $\alpha, \beta, \ldots$ & run from $0$ to $3$\\
  \hline
  Latin lowercase indices $i, j, \ldots$ & run from $1$ to $8$\\
  \hline
  $\mathbf G$, $G^i$ & geodesic spray and its component\\
  \hline
  $\pi_M$ & projection map from $TM$ to $M$\\
  \hline
  $\gamma$, $\gamma^\mu$ & geodesic curve and its representation in a coordinate system\\
  \hline
  $\dot \gamma$ & ordinary derivative of $\gamma$ wrt the affine parameter\\
  \hline
  $\phi$ & coordinate function of a coordinate chart; geodesic flow\\
  \hline
  $\wedge$ & wedge product\\
  \hline
  $\iprod$ & interior product\\
  \hline
  $\calL_{\mathbf X}$ & Lie derivative along an element of tangent space\\
  \hline
  $\cong$ & is isomorphic to\\
  \hline
  $\oplus$ & direct sum\\
  \hline
  subscripts $H$, $V$ & a part of a vector fully contained in horizontal or vertical subspace\\
  \hline
  superscript $T$ & generalized transpose of operator\\
  \hline
  
\end{tabular}
\egroup

\tableofcontents

\chapter{General introduction}
\pagenumbering{arabic}

The theory of transfer matrices and propagators is a fairly 
common tool in
theoretical physics and mathematics. In many physical systems the propagation of disturbances can be described as 
the propagation of waves, which at some level of approximation can be described using resolvent operators mapping some initial state of the
physical system to the state at some later time. This approach has been used in many branches of physics ranging from the theory of elasticity
to the quantum field theory.  

In astrophysics and cosmology, most observations require a good understanding of the propagation of electromagnetic waves through a curved spacetime.  
In the high-frequency limit the propagation of waves, which is almost always applicable in astrophysical situations,
  can be well approximated by propagation along null geodesics \cite{ehlers1967,isaacson1968,isaacson1968b,misner1973,perlick2000,harte2019}. Still, for sources of
a finite extent we need to consider a whole family of null geodesics, corresponding to the light rays from different points of the source's cross-section. While the general problem of light ray propagation is non-linear, it can be simplified for geodesics remaining close to a given one. Namely, if one geodesic is known, then the behaviour of neighbouring null geodesics can be described sufficiently well by the first order
 \emph{geodesic 
 deviation equation} (GDE). As a system of linear ordinary differential equations, it admits the propagation matrix 
 or resolvent formulation, in which the solution at a later time can be obtained from the state at an earlier moment
 by the action of a linear operator.
 
 In the most general sense, the geodesic deviation equation relates the relative motions of physical particles or light rays to the spacetime geometry. However,
 to our knowledge, 
 there has been still no complete discussion in the literature of the transfer matrix technique
 to the geodesic deviation equation in the case of null geodesics, representing light rays. The goal of this 
 thesis is to fill in this gap and provide a new perspective on geometric optics in general relativity. 
 Following our previous work, we will refer to this framework as  the \emph{bilocal geodesic
 operator}, or 
 BGO formalism. 

The BGO formalism can describe all nontrivial optical
effects as experienced by a source and an observer in two regions of spacetime connected by a null geodesic. The effects are 
considered at the lowest
 order in the perturbations of the
positions of the source and the observer. The main starting point is the GDE around a null geodesic. Note that this equation is
 fully relativistic,
in the sense that it makes no approximations regarding the metric tensor, such as the Newtonian or post-Newtonian approximation. 
On the other hand, unlike the more familiar optical scalars formalism (also known as Sachs formalism) \cite{sachs1961,perlick2004} it is based on a
 linear system of equations. We show in this work that this opens up the possibility to apply directly the machinery of linear algebra to many problems in geometrical optics.
 
 The formalism can describe, among other things, the standard effects of gravitational lensing, in the form of the
 magnification and elliptical deformation of the source's image. This information is stored in a lower-dimensional operator known as the Jacobi matrix \cite{schneider1992,perlick2004}\footnote{Here by the Jacobi matrix $D_{ij}$ we mean the resolvent operator of the GDE projected on the screen space. In the weak lensing literature there is a similar but different Jacobian matrix of the lens equation (also known as the amplification or magnification matrix \cite{schneider1992}) defined by $A_{ij} = \dfrac{\partial \beta^i}{\partial \theta^j}$. The crucial difference is that the second matrix requires a clear separation of the background and the lens. In general, they are related by $D_{ij}={\paren{D_B}}_{ik} A_{kj}$ with $D_B$ being the solution on the background, but in practice for homogeneous and isotropic spacetimes $\paren {D_B}_{ij}$ can be replaced by an averaged area distance, i.e. $\bar d_A \delta_{ij}$\cite{clarkson2015}.}. However, in its most general form it also includes geodesic perturbations 
 related to the variation of the observer's position or the observation time. This way the BGO formalism expands the standard set of observables by including the 
 effects of the observer's displacements, i.e the parallax
 effects, as well as the time variations of the redshift and position, i.e. the drift effects. Furthermore, in the BGO formalism,
 it is relatively easy to show that 
  these effects are in fact related to each other and derive precise mathematical relations between the observables.
  As an example, the second paper from this thesis \cite{korzynski2022}
  explores
  general relations between the magnification of a source and the parallax effects.

Let us mention that the mathematical apparatus of propagator matrices in GR has been considered earlier
along both timelike and null geodesics.
 In the timelike case, it was used in the studies of the motion of massive extended \cite{dixon1970} and charged \cite{dewitt1960} bodies (although these applications
 differ from ours, because in these papers propagator matrices were used to define covariant averages and energy exchange domains rather than to track
 the motion of a family of non-interacting particles or light rays), also the singularity theorems \cite{hawking1966,hawking2014, hawking2023} and 
 the gravitational wave memory effects \cite{flanagan2019} under the name of Jacobi propagator\footnote{In the mathematical literature of Lorentzian manifolds it is known as the Jacobi tensor \cite{beem1996global}}. 
 In the null case the resolvent formalism was used to study and gravitational radiation \cite{newman1962} and general relativistic energy flux \cite{penrose1966fl}.
 In the cosmological setting, the formalism was occasionally used as a technical tool to describe gravitational lensing  in inhomogeneous Universe models \cite{fleury2014}. However, the authors restrict the space of solutions of the GDE to 
 those which correspond to momentary observations, excluding this way any drift effects.

\marginnote{Updated paragraph} A different treatment of propagation of light rests upon Synge's world 
function \cite{ruse1931, synge1960, vines2015, korzynski2021} which has also found applications in relativistic
self-force \cite{poisson2011, barack2018} and extended body problems \cite{dixon1970, harte2012m}. Namely, it turns out that the first and second 
derivatives of the world function have a direct connection
 to the solutions of the geodesic equation and the GDE respectively. This approach is less useful in practice since the formula for the world function requires full 
 knowledge of the solution to the GE, which in general is very difficult. Therefore, we do not follow this approach in this Thesis, although 
 we note here that the BGO formalism can indeed be formulated in the language of the world function \cite{korzynski2021}. 

 \marginnote{Addendum}Still, there are a few interesting applications worth pointing out. In theoretical studies, Synge function was 
 used to study caustics and optical properties of nonlinear gravitational waves \cite{harte2012,harte2013}, 
 local surface of communication \cite{korzynski2021}. Moreover, a closely related 
 van Vleck determinant was again used in \cite{harte2012} but also to investigate geodesic focusing \cite{visser1993},
 optical distance measures \cite{ivanov2018, ivanov2018p} and gravitational magnification \cite{werner2015,aazami2016}.

 \marginnote{Addendum}In practice, the world function serves as a precursor to the time transfer
function \cite{linet2002, le2004} which allows the determination of the duration of the propagation
of a light signal when the coordinates of both endpoints are known. It is
very convenient for the perturbative description of the problem.

As mentioned above, the wave-like propagation of disturbances and their high-frequency
 limit can also be found in other contexts, for example, in deformable media which falls under the theory of seismology. In this context,
  the solutions of the elastodynamic equations in high-frequency regime can be approximated by seismic rays, 
  and their propagation in an appropriate approximation is ruled by bilocal operators, quite similar to the BGOs 
   \cite{farra1999,cerveny2001,cerveny2012,waheed2013}.

More recently, the BGO approach has also appeared in other works in a slightly 
different formulation. In \cite{gallo2011, crisnejo2018} the authors applied the perturbative expansion of the solution 
to the GDE in powers of the Riemann tensor, and later used it to estimate the optical scalars. In \cite{uzun2020}, the non-relativistic
 counterpart of BGOs was presented, together with an extensive discussion of its symplectic properties. However, the authors again limit their considerations to momentary observations and do not 
 consider the drift effects.

The aim of the first part of this thesis is to provide a solid geometrical foundation for the BGO formalism
 valid for any smooth manifold equipped with a metric and its Levi-Civita connection. Then we will use this to display the relation
  of BGOs to various optical effects and observables.
    
  The two papers included in this Thesis illustrate the application of the BGOs. In the first paper we consider the BGOs in static spherically symmetric spacetimes which are derived in two different ways. The first approach one to vary the solution to the GE with respect to its initial conditions. However, a simple variation leads to non-covariant results. The problem is averted by making sure the variations themselves are expressed in a covariant language. The geometric interpretation of this requirement will be outlined in the mathematical introduction to the Thesis, in Sec. \ref{sec:geomtb}. The second method we propose directly involves the GDE, its conservation laws, and symplectic properties, whose geometric foundations will be laid in Secs. \ref{sec:geodspr}-\ref{sec:symp}. Finally, the BGOs are used to describe optical distance measures in Schwarzschild spacetime.

In the second paper, we prove several distance inequalities assuming GR, propagation of light in vacuum, and reasonable conditions on the matter. Inequalities themselves are proved with the use of Sachs optical equations \cite{sachs1961}, while the BGOs allow us to relate the parallax effect to the behaviour of a special bundle of rays. This relation is crucial for the proof of inequality.

We would also like to bring attention to a complementary work on BGOs \cite{grasso2021a,grasso2021b,grasso2021c}. In these papers, the authors introduce a Mathematica package to calculate the BGOs in arbitrary evolving spacetimes. Similarly to our approach here, the authors rewrite the GDE as two systems of the first order ODEs which are then integrated backward in time. Then, to integrate the solution forward in time they apply the symplecticity of the total resolvent operator. One of the goals of this thesis is to reveal a deeper geometrical meaning of the first-order formulation and explain why the total transfer matrix is not just a formal construction.

\chapter{Mathematical preliminaries}
\label{mathin}
\section{Introduction}

The \emph{bilocal geodesic operators}, or BGOs, are a very useful tool in the study of light propagation between distant regions of spacetime. 
They relate deviations of the initial point and its tangent vector to their counterparts at the other end of the curve. Since they are defined for arbitrary initial data sets, they describe how a curved spacetime influences the evolution of any perturbed geodesic, provided that the perturbation is not too large. In the context of light propagation, the BGOs describe how a perturbed light ray is bent by the spacetime curvature, or, more broadly, how any whole infinitesimal bundle of light rays is affected by the curvature. They also describe how the results of observations performed at the observer's endpoint of the geodesic segment vary with respect to the observer's proper time. In the relativistic and cosmological literature these variations are known as drifts.
Physical aspects of drift effects \cite{korzynski2018}, general relativistic parallax \cite{korzynski2022}, times of arrivals \cite{korzynski2021}, and BGOs \cite{grasso2019} have been discussed previously, along with some geometrical remarks. 

This chapter, apart from serving as a mathematical introduction, contains also the first part of the new results of the Thesis. We show here how we can define the BGOs with the help of the geometry of the tangent bundle. We will therefore begin by briefly reviewing the basic notions of differential geometry, such as the tangent bundle, geodesics, and their lifts to the tangent bundle. Recall that the geodesics induce a special vector field on the tangent bundle known as the geodesic spray, which in turn defines the geodesic flow. This flow is analogous to a fluid flow in the mechanics of continuous media. We will show that the tangent map to this flow, or the deformation gradient tensor of the fictitious fluid related to the flow,  naturally splits into $4$ covariant, bilocal operators we can identify with the BGOs.

In the meantime, we also develop a geometrical formulation of the techniques used in the first paper \cite{serbenta2022} contained in this Thesis, in which we present two methods of solving the GDE exactly and determining the BGOs. One of them requires the computation of the total variation of a geodesic. However, the variation needs to be decomposed in a covariant manner into the position variation at a point and the variation of the tangent vector. As we will see, the meaning of this step is easy to understand if we express it in the language of the tangent bundle. The key construction in this case is the covariant splitting of the tangent space to the tangent bundle into the horizontal and vertical subspaces. 

The geodesic flow preserves the symplectic structure of the tangent bundle \cite{uzun2020,kijowski1979}. It turns out that this has a direct implication for the symplectic properties of BGOs and light propagation, see for example the proof of the distance inequality \cite{korzynski2022}. Therefore we will also discuss the relation between the BGOs and the symplectic geometry of the tangent bundle in the later part of this section.
 
As a general introduction to the topics presented below, we recommend the following references  \cite{lindquist1966, yano1973, besse1978, lang1985, sakai1996, lang2012, ilmavirta2020}, while more specialized references are given in the text.

\section{Charts and projections}
\label{sec:chpr}

The purpose of this section is to introduce some basic elements of differential geometry on which we will build the formalism for BGOs. We begin with the basic concepts of coordinates and tangent spaces on a manifold and then show how they apply to the tangent bundle. Since the main goal of this work is to 
understand the physics of light propagation,  we will also discuss the geometric and physical interpretation of the elements of tangent spaces 
in the context of geometric optics in curved spacetimes.

We begin by setting up the model for spacetime.  The spacetime will be represented by a smooth 4-dimensional Lorentzian manifold $M$ with metric $g$ and its Levi-Civita connection $\nabla$. For its signature we choose $(-+++)$. We note that although we are interested in manifolds describing the spacetime geometry, the theory to be presented below can be equally applied to Riemannian or semi-Riemannian manifolds of arbitrary finite dimensionality.

Every element of $M$ is simply a point, or more precisely, an event in spacetime. In order to assign coordinates to this point we need to introduce a \textit{chart}. By chart we mean the structure $\tuple{U, \phi, \R^4}$, where $U \subset M$ is an open set and $\phi$ is a map from $U$ to $\R^4$. If $(x^\mu)$ is our coordinate system on $U$, then
\begin{equation}
\forall p \in U : \map \phi p = (\map {x^\mu} p).
\end{equation}
That is, $\phi$ allows us to assign to each point of the spacetime 4 numbers $x^\mu$ known as coordinates, in a smooth way. Note that from the point of view of differential geometry $x^\mu$ does not transform as a component of a vector. For the sake of simplicity, we assume the manifold $M$ and the metric $g$ to be smooth.

Tangent vectors to a manifold at a point form a vector space known as the \textit{tangent space}. While there are several ways to introduce it, we will follow here the standard approach, resting upon linear maps $X : \map {C^k} M \to \R$ acting on functions, which possess Leibniz-like properties of differentiation and are known as \textit{derivations} \cite{sakai1996,lee2013}. At any $p \in M$ the set of derivations forms a vector space with a well-defined natural basis $\set {\valat{\dfrac{\partial}{\partial x^\mu}} p}$, given by 
partial derivatives with respect to all coordinates, known as the \textit{coordinate basis}. Any so-constructed vector space at $p\in M$ is denoted by $T_p M$ and called \emph{the tangent space at $p$}. 

The formal definition using linear operators is somewhat too abstract
for our purposes, therefore we will also consider its other geometric interpretations. 
First, consider all differentiable curves passing through a point $p \in M$. Then all their tangent vectors at $p$ span a vector space at this point. A more physical way of thinking about tangent vectors is to depict them as all possible small displacements of the point $p$. We can think either of infinitesimal displacement or, alternatively, of small but finite displacements, such that the scale of the displacement is much smaller than the curvature scale or any other relevant length scale in the neighbourhood of $p$. The latter interpretation requires the use of Riemann's normal coordinates,
see \cite{wald2010, lee2018}.  In other words, a vector in $T_p M$ can also stand for a perturbation of the position of an object in the spacetime or a more general manifold. In this Thesis we will make extensive use of this interpretation
since it fits best the physical situations we want to describe.

In principle, the manifold and its tangent spaces provide a sufficient background to study problems describable by 2nd order ordinary differential equations (ODEs), such as the GDE. However, the same problems can be recast as a 1st order ODE in a higher-dimensional setting \cite{jerie2002}. The order reduction is advantageous because it allows us to study all degrees of freedom on equal footing and apply the resolvent formalism to the equation. In the case of the geodesic equation and the geodesic deviation equation this can be achieved by considering the whole \textit{tangent bundle} $TM$ \cite{sasaki1958,sasaki1962,dombrowski1962, gudmundsson2002, bucataru2007,yano1968}, which is a disjoint union of all tangent spaces:

\begin{equation}
TM = \bigsqcup_{p \mathop \in M} T_p M = \set {\tuple {p, \mathbf X_p} : p \in M, \mathbf X_p \in T_p M}.
\end{equation}
It comes with a natural \textit{projection map} $\pi_M$ satisfying
\begin{equation}
\pi_M : TM \ni \tuple {p, \mathbf X_p} \to p \in M.
\end{equation}
One important property of the tangent bundle is that it is also a manifold. Therefore, we can  introduce on it geometrical objects such as charts, coordinates, curves and vector fields.

 While there are many possible charts on $TM$, there is a class of natural charts induced by the charts in the base manifold $M$. They can be considered as follows. Let $\tilde U$ be an open set in $TM$ defined by $\tilde U = \map {\pi_M^{-1}} U$, where $U$ is an open set in $M$ defining a chart. Suppose the coordinates of $p \in U$ read $\tuple {x^\mu}$. Suppose also that a vector $\mathbf X_p \in T_p M$ is tangent to $M$ at $p$. A vector can be expanded with respect to the local basis as  $\ds \mathbf X_p = v^\mu \valat {\frac{\partial}{\partial x^\mu}} p$, where $v^\mu$ are the components of $X_p$ in the coordinate basis. Let $\tilde \phi$ be a map from $\tilde U$ to $\R^8$ such that
\begin{equation}
\forall \tuple {p, \mathbf X_p} \in \tilde U : \map{\tilde \phi} {\tuple {p, \mathbf X_p}} = \tuple {\tilde x^\mu, v^\nu},
\end{equation}
where
\begin{equation}
\map {\pi_M} {\tilde x^\mu, v^\nu} = x^\mu,
\end{equation}
or equivalently
\begin{equation}
\map {\tilde x^\mu} {p, \mathbf X_p} = \map {x^\mu} p.
\end{equation}
Then the structure $\struct {\tilde U, \tilde \phi, \R^8}$ defines a (smooth) chart on $TM$ and $\tuple {\chi^i} = \tuple {\tilde x^\mu, v^\nu}$ are the local coordinates. Note that $x^\mu$ and $\tilde x^\mu$ are images of different mappings and so cannot be identical, but in practice their meaning is synonymous. Therefore, from now on we will drop the tilde and denote the coordinates on $TM$ as $\tuple {x^\mu, v^\mu}$. These coordinates are known as the \emph{adapted} \cite{sarbach2014g}, \emph{induced} \cite{yano1973}, or \emph{canonically associated} \cite{besse1978} coordinates. The dimension of the tangent bundle is twice the dimension of the base manifold, so in our case it is 8.

In comparison with $M$, its tangent bundle admits a much greater variety of curves, simply because of its higher dimension: for every curve on the base manifold there is an entire family of curves related to the former by the projection $\pi_M$. However, among all the curves in the family corresponding to a given curve in $M$ there is a unique one known as the \textit{lift}, which is defined as follows. Consider a differentiable curve on the spacetime $\gamma : \R \supseteq I \to M$. Let $\Gamma : \R \supseteq I \to TM$ be a curve
in the tangent bundle such that for all $\lambda \in I$ the curve $\Gamma$ satisfies $\map \Gamma \lambda = \tuple {\map \gamma \lambda, \map {\dot \gamma} \lambda}$, that is, which in local coordinates of $TM$ reads
\begin{equation}
\begin{cases}
\map {x^\mu} \lambda &= \map {\gamma^\mu} \lambda\\
\map {v^\mu} \lambda &= \map {\dot \gamma^\mu} \lambda.
\end{cases}
\end{equation}
Then we call $\Gamma$ the lift of $\gamma$. Obviously, $\map {\pi_M} \Gamma = \gamma$. The main benefit of lifting a curve to $TM$ is that the curves passing through the same point in $M$ but differing by their tangent vectors do not cross on $TM$.

Finally, since our goal is the description of the perturbations of curves, we need to introduce the tangent space to $TM$. Pick any point $\tuple {p, \mathbf X_p}$ in $TM$. Since $TM$ is a manifold, at every point it possesses a tangent space that we will denote by $T_{\tuple {p, \mathbf X_p}} TM$. Again, as a vector space, it has basis, and the coordinates $\tuple {x^\mu, v^\nu}$ induce the \textit{associated basis} $\set {\mathbf f_{x, \mu}, \mathbf f_{v, \mu}}$, where
\begin{equation}
\label{eq:trb}
\begin{split}
 \mathbf f_{x, \mu} & =\valat {\ds \frac{\partial}{\partial x^\mu}} {\tuple {p, \mathbf X_p}}\\
 \mathbf f_{v, \mu} &= \valat {\ds \frac{\partial}{\partial v^\nu}} {\tuple {p, \mathbf X_p}},
 \end{split}
\end{equation}
i.e. the basis vectors are given by the partial derivatives with all other coordinates fixed. 
In analogy with the previous case, the elements of $T_{\tuple {p, \mathbf X_p}} TM$ can be understood as either tangent vectors to curves in $TM$ passing through $\tuple {p, \mathbf X_p}$ or as infinitesimal perturbations of the point $\tuple {p, \mathbf X_p}$. In the latter case we may think of it collectively as the perturbation of both the point $p \in M$ \emph{and at the same time}, of the vector $\mathbf X$ tangent to $M$ at $p$.

\section{Geodesic flow}
\label{sec:geofl}

Geodesics are arguably the most important family of curves in 
general relativity. They represent the worldlines of massive and massless
particles interacting only with the gravitational field.
While geodesics themselves are curves on the base manifold, it is 
often useful to consider them on the level of the tangent bundle. 
The main advantage of this approach is that the geodesic motion in the 
tangent bundle can be represented by a 
flow of the whole tangent bundle along a vector field, called the \emph{geodesic flow}. 

The geodesic flow is the central notion of this 
chapter, because, as we will see later, the BGOs can be defined 
in terms of its properties. Namely, the BGOs are related to 
its action on infinitesimal volume elements, or, more precisely, 
to the deformation 
gradient of the geodesic flow. Therefore  we will review in this chapter 
its 
definition and discuss some of its fundamental properties.

In general relativity, a freely falling particle follows a 
uniquely defined curve known as the \textit{geodesic}. It is a smooth 
curve $\gamma : \R \supset I \to M$ such 
 that its tangent vector is parallel transported along itself:
\begin{equation}
\label{eq:gea}
\nabla_{\dot \gamma}\dot \gamma = 0.
\end{equation}
In coordinates this equation reads
\begin{equation}
\frac{d^2 \gamma^\mu}{d \lambda^2} + \Gamma^\mu_{~\alpha\beta}\frac{d \gamma^\alpha}{d \lambda} \frac{d \gamma^\beta}{d \lambda} = 0,
\end{equation}
where $\gamma^\mu = x^\mu \circ \gamma$ and $\displaystyle \dot \gamma^\mu = \frac{d \gamma^\mu}{d \lambda}$. Moreover, we also assume that the geodesic depends smoothly on its initial data. This property will be utilised later to find solutions of the GDE.

Usually, geodesics are treated as curves in the base manifold $M$. 
However, we can also consider closely related curves generated by 
lifting the geodesics to $TM$. It turns out that this construction 
is closely related to the Hamiltonian approach to the geodesic motion.
 More precisely, the phase space for geodesic motion is naturally 
 modeled by the cotangent bundle $T^* M$.  In a manifold without
 additional structures $T^* M$ is unrelated 
 to $TM$. However, the metric $g$ provides a natural isomorphism 
 between both spaces, so $TM$ and $T^*M$ can be considered interchangeably. 

A particularly useful representation of all the geodesics on the 
tangent bundle is known as the geodesic flow \cite{paternain2012}. 
It is a family of diffeomorphisms $\phi_\lambda : TM \times \R \to TM$
 of the tangent bundle defined by
\begin{equation}
\label{eq:gf}
    \map {\phi_\lambda} {p, \mathbf X_p} = \tuple {\map {\gamma_{\tuple {p, \mathbf X_p}}} \lambda, \map {\dot \gamma_{\tuple {p, \mathbf X_p}}} \lambda}
\end{equation}
where $\gamma$ and $\dot \gamma$ is the geodesic and its 
derivative with respect to $\lambda$ with initial 
conditions $\map \gamma {\lambda_0} = p$ and $\map {\dot \gamma} {\lambda_0} = \mathbf X_p$. 

The fact that $\phi_\lambda$ is a flow can be seen by checking
 that is satisfies $\phi_{\lambda_1 + \lambda_2} = \phi_{\lambda_1} \circ \phi_{\lambda_2}$. 
 Also, since for geodesics the length of their tangent vectors are 
 conserved, it follows that the geodesic flow preserves a number of 
 subbundles of $TM$. Namely, let $S^+TM$, $S^-TM$ and $S^0TM$ be 
 subbundles of $TM$ in which the geodesics respectively satisfy the 
 constraints $\dot \gamma^\mu \dot \gamma_\mu = 1$, $\dot \gamma^\mu \dot \gamma_\mu = -1$ and 
 $\dot \gamma^\mu \dot \gamma_\mu = 0$. Then $S^+TM$, $S^-TM$ and 
 $S^0TM$ are preserved by the flow $\phi_\lambda$. 

\section{Vertical and horizontal subspaces}
\label{sec:geomtb}

 The dimension of the tangent bundle is twice the dimension of the 
 spacetime. Moreover, the tangent space of the tangent bundle 
 has an associated basis \eqref{eq:trb}, whose form suggests the 
 possibility of splitting any vector into two independent parts 
proportional to vectors $\partial/\partial x^\mu$ and $\partial/\partial v^\mu$ respectively. However, this 
 kind of splitting is not covariant, i.e. it depends on the adapted 
 coordinates we have chosen. Nevertheless, in the case of spacetime endowed 
 with a non-degenerate metric and its Levi-Civita connection, we have additional structure. As we will show in this section, this leads to an invariant splitting of 
  $T_{\tuple {p, \mathbf X_p}}TM$ into two $4$-dimensional spaces, 
  both isomorphic to $T_p M$ \cite{gliklikh2011}. In other words. the construction is coordinate 
  system-independent and permits the representation of 
  vectors in $T_{\tuple {p, \mathbf X_p}}TM$ as pairs of tangent
  vectors to the spacetime. 
  This splitting is crucial for 
  the definition of the BGOs, so in this section we will describe it in detail and discuss its physical interpretation as well as its relation to the observations in geometrical
   optics.

Previously we have shown that the tangent bundle $TM$ has a natural
 coordinate system $\tuple {x^\mu, v^\mu}$, while its tangent 
 space $T_{\tuple{p, \mathbf X_p}} TM$ can be equipped with the 
 basis $\set {\mathbf f_{x,\mu}, \mathbf f_{v,\mu} }$ defined by 
 the partial derivatives with respect to the coordinates,
 see (\ref{eq:trb}).
In the associated basis the expansion of a vector 
$\mathbf X \in T_{\tuple{p, \mathbf X_p}} TM$ reads
\begin{equation}
\label{eq:dttm}
   \mathbf X = X^\mu_x \dfrac{\partial}{\partial x^\mu} + X^\mu_{v} \dfrac{\partial}{\partial v^\mu}.
\end{equation}
Vector $\mathbf X$ corresponds to a perturbation of
 a point in $TM$, i.e. a perturbation of a point in $M$ and a vector 
 tangent at that point. In the decomposition above $X^\mu_x$ stands obviously
 for the variation of 
 position in $M$, and $X^\mu_v$ for the variation of the vector. 
 This decomposition, however, is not covariant. 
 Consider a general transformation of adapted coordinates on $TM$, 
 induced by coordinates on $M$. Then the 
 transformation from $\tuple {x^\mu, v^\mu}$ to
  $\tuple {\tilde x^{\tilde \mu}, \tilde v^{\tilde \mu}}$ reads
\begin{equation}
\label{eq:indtr}
\begin{cases}
\begin{split}
\tilde x^{\tilde \mu} &= \map {\tilde x^{\tilde \mu}} {x^\mu}\\
\tilde v^{\tilde \nu} &= v^\nu \frac{\partial \tilde x^{\tilde \nu}}{\partial x^\nu},
\end{split}
\end{cases}
\end{equation}
where $\dfrac {\partial \tilde x^{\tilde \nu}}{\partial x^\nu}$ is the
 Jacobian matrix of the coordinate transformation on $M$.
It follows that the associated basis vectors transform according to
\begin{equation}
\begin{cases}
\begin{split}
\frac{\partial}{\partial \tilde v^{\tilde \nu}} &= \frac{\partial x^\nu}{\partial \tilde x^{\tilde \nu}} \frac{\partial}{\partial v^\nu}\\
\frac{\partial}{\partial \tilde x^{\tilde \mu}} & = \frac{\partial v^\nu}{\partial \tilde x^{\tilde \mu}} \frac{\partial}{\partial v^\nu} + \frac{\partial x^\mu}{\partial \tilde x^{\tilde \mu} } \frac{\partial}{\partial x^\mu}.
\end{split}
\end{cases}
\end{equation} 
This in turn implies that the Jacobian matrix
of the coordinate transform in $TM$ reads\footnote{We remind that for a given coordinate system on the tagent bundle $x^\mu$ and $v^\mu$  are considered independent of each other.}
\begin{equation}
\frac{\partial \tuple{\tilde x^{\tilde \mu}, \tilde v^{\tilde \mu}}}{\partial \tuple{x^\nu, v^\nu}} = \paren{
\begin{array}{cc}
\dfrac{\partial \tilde x^{\tilde \mu}}{\partial x^\nu} &  0\\[2ex]
 v^\rho \dfrac{\partial^2 \tilde x^{\tilde \mu}}{\partial x^\nu \partial x^\rho} &  \dfrac{\partial \tilde x^{\tilde \mu}}{\partial x^\nu}\end{array}}.
\end{equation}
From the structure of the transformations it follows that the decomposition \eqref{eq:dttm} is not covariant. Namely, after a general coordinate transformation
 on $TM$ the components proportional to $\partial/\partial \tilde x^{\tilde \mu}$ may generate terms proportional to 
 $\partial/\partial v^\mu$. However, we can easily
  identify one invariant subspace $V_{\tuple{p, \mathbf X_p}} \subset T_{\tuple{p, \mathbf X_p}} TM$,
   defined by
\begin{equation}
V_{\tuple{p, \mathbf X_p}} = \set {\mathbf X \in T_{\tuple {p, \mathbf X_p}}TM : \mathbf X = \map {\operatorname{span}} {\frac{\partial}{\partial v^\alpha}} }
\end{equation}
that we will call the \textit{vertical subspace}. It is invariant in the sense that the subspace defined this way does not depend on the associated basis. The definition of  $V$ can also be stated in the language of the 
projection maps. Let
\begin{equation}
d_{\tuple {p, \mathbf X_p}} \pi_M : T_{\tuple {p, \mathbf X_p}}TM \to T_p M
\end{equation}
 denote the differential (or the pushforward map) of the
 projection $\pi_M$. Its action on the basis vectors yields
\begin{equation}
\begin{split}
\map {d_{\tuple{p, \mathbf X_p}} \pi_M} {\valat {\frac{\partial}{\partial x^\mu}} {\tuple {p, \mathbf X_p}}} &= \valat {\frac{\partial}{\partial x^\mu}} p\\
\map {d_{\tuple{p, \mathbf X_p}} \pi_M} {\valat {\frac{\partial}{\partial v^\mu}} {\tuple {p, \mathbf X_p}}} &= 0.
\end{split}
\end{equation} 
Hence, $\mathbf X$ belongs to the vertical subspace 
if $\map {d_{\tuple {p, \mathbf X_p}} \pi_M} {\mathbf X} = 0$, 
and we can write $V_{\tuple {p, \mathbf X_p}} = 
\map {\operatorname {ker}} {d_{\tuple {p, \mathbf X_p}} \pi_M}$. 
In other words, a vector in the tangent space to the 
tangent bundle belongs to the  
vertical subspace if the corresponding the infinitesimal 
variation of the base point in $M$ vanishes.

A complementary subspace to $V$ can be constructed for
 manifolds endowed with an additional structure like the metric 
 and the connection. As we have noted before, a perturbation of the 
 position after a change of coordinates may acquire 
 a component along $\partial/\partial v^\nu$.
  Hence, a general perturbation of the position does not preserve the decomposition with respect to the coordinate basis on $TM$.
  However, the existence of the Levi-Civita connection 
  suggests that a slight modification of the basis motivated by the parallel transport of vectors on $T_p M$ might avoid mixing the terms. 
  To check if this is the case, let us rewrite \eqref{eq:dttm} by
   adding and subtracting same terms:
\begin{equation}
\mathbf X = X^\mu_x \frac{\partial}{\partial x^\mu} + X^\mu_v \frac{\partial}{\partial v^\mu} = X^\mu_x \paren {\frac{\partial}{\partial x^\mu} - \Gamma^\alpha_{~\mu\beta} v^\beta \frac{\partial}{\partial v^\alpha}} + \paren {X^\mu_v + \Gamma^\mu_{~\alpha\beta} X^\alpha_x v^\beta}\frac{\partial}{\partial v^\mu}.
\end{equation}
We now introduce the following notation:
\begin{equation}
\mathbf X = X^\mu_H \mathbf e_{H\mu} + X^\mu_V \mathbf e_{V\mu},
\end{equation}
where
\begin{equation}
\label{eq:hvid}
\begin{split}
X^\mu_H &= X^\mu_x\\
X^\mu_V &= X^\mu_v + \Gamma^\mu_{~\alpha\beta} X^\alpha_x v^\beta,
\end{split}
\end{equation}
and the new basis reads:
\begin{equation}
\begin{split}
\mathbf e_{H\mu} &= \frac{\partial}{\partial x^\mu} - \Gamma^\alpha_{~\mu\beta}v^\beta \frac{\partial}{\partial v^\alpha}\\
\mathbf e_{V\mu} &= \frac{\partial}{\partial v^\mu}.
\end{split}
\end{equation}
Again, the part proportional to $\mathbf e_V$ belongs to $V_{\tuple {p, \mathbf X_p}}$. However, it is easy to check that coordinate transformations \eqref{eq:indtr} do not mix $\mathbf e_H$ and $\mathbf e_V$. Therefore, by setting $X^\mu_V = 0$ we obtain another invariant subspace of $T_{\tuple{p, \mathbf X_p}}TM$ \cite{sarbach2013, sarbach2014g, sarbach2014}, which we call the \textit{horizontal subspace} and denote by $H_{\tuple{p, \mathbf X_p}}$:
\begin{equation}
H_{\tuple{p, \mathbf X_p}} = \set{\mathbf X \in T_{\tuple{p, \mathbf X_p}}TM : \mathbf X = \map {\operatorname {span}} {\mathbf e_H}}.
\end{equation}
The condition above is equivalent to the requirement that
 $X^\mu_v + \Gamma^\mu_{~\alpha\beta} X^\alpha_x v^\beta = 0$.
 In geometric terms, the equation above means that
  that the tangent vector $\mathbf X_p$ is parallel transported
  when the base point $p$ varies.

The horizontal and vertical subspaces as presented above are 
invariant under the choice of coordinates on the base manifold, 
complementary and sufficient to span the entire $T_{\tuple {p, \mathbf X_p}} TM$. 
That is, $T_{\tuple {p, \mathbf X_p}} TM = V_{\tuple {p, \mathbf X_p}} 
\oplus H_{\tuple {p, \mathbf X_p}}$ and $V_{\tuple {p, \mathbf X_p}} 
\cap H_{\tuple {p, \mathbf X_p}} = \set 0$. 
Hence, on the tangent space of $TM$ any vector can be decomposed 
uniquely into horizontal and vertical parts. 

Both horizontal and vertical subspaces are isomorphic to the 
tangent space $T_p M$, i.e. there are coordinate-independent isomorphisms
 $H_{\tuple {p, \mathbf X_p}} \cong T_p M$ and 
 $V_{\tuple {p, \mathbf X_p}} \cong T_p M$ \cite{besse1978,sakai1996}.
 In terms of coordinates they are given by
 \begin{equation}
 \label{eq:iso}
    \begin{split}
       \operatorname{iso}^H &: H_{\tuple {p, \mathbf X_p}} \ni X_H^\mu \, {\mathbf e}_{H\mu}
        \mapsto X_H^\mu \, \left.\frac{\partial}{\partial x^\mu}\right|_p \in T_p M\\
       \operatorname{iso}^V &: V_{\tuple {p, \mathbf X_p}} \ni X_V^\mu \, {\mathbf e}_{V\mu}
        \mapsto X_V^\mu \, \left.\frac{\partial}{\partial x^\mu}\right|_p \in T_p M
    \end{split}
\end{equation}

In simple terms, we have shown that any infinitesimal variation 
of a point and a vector at that point can be covariantly decomposed 
into two components. The first one, i.e. the horizontal component, 
corresponds to the variation of the point together with the parallel 
transport of the vector along this variation. The second one, i.e. 
the vertical part is simply the variation of the vector without 
changing its base point. Both components
can be parametrized by vectors from the tangent space to 
the spacetime $T_p M$. In other words, an infinitesimal 
variation of a point and a vector can be uniquely 
represented by a \emph{pair} of vectors tangent to the spacetime,
each representing a different type of variation. 

The horizontal-vertical splitting presented above is particularly 
useful in the context of geometrical optics,
because the observables, such as the position drift or parallax, are 
directly related
to the covariant variation of the direction of light propagation
between nearby points of the spacetime \cite{grasso2019}.
The covariant direction variation, on the other hand, is
given by the vertical components of the variation of the 
observation point and the light propagation direction.

\section{Geodesic spray}
\label{sec:geodspr}

We know that for a curve to be geodesic on $M$, its tangent vector 
has to satisfy Eq. \eqref{eq:gea}. Suppose we are given \emph{a non-vanishing}
 vector field $\dot \gamma$ satisfying the geodesic condition and
  defined on the whole manifold. The 3-parameter family of its integral 
  curves is known as the geodesic congruence. In a similar way, 
  we can define a congruence on $TM$ that consists of 
  lifts of all possible geodesics on the spacetime. The vector field
   on $TM$ which generates these curves (in fact, the entire geodesic flow introduced in Sec. \ref{sec:geofl}) is known as the \textit{geodesic spray}. In other
     words, the geodesic spray is an infinitesimal counterpart of the
      geodesic flow on the level of $TM$. In this section 
we will recall the notion of the geodesic spray and review its geometric properties. In 
particular, we will show that the geodesic spray preserves the
 symplectic structure, which allows us to study the geodesic motion 
 in the Hamiltonian formulation. In later sections dedicated to null geodesics, these geometrical features will be related to the physical properties of light rays and their propagation.

The decomposition of $T_{\tuple {p, \mathbf X_p}} TM$ we have introduced
previously holds 
for any vector field on $TM$. In our work we apply it to a specific 
vector field,
namely the geodesic spray. It is defined
as a vector field $\mathbf G_{\tuple {p, \mathbf X_p}} : 
TM \ni \tuple{p, \mathbf X_p} \to T_{\tuple{p, \mathbf X_p}} TM$ 
\cite{crampin1984}, which in adapted coordinates
 reads
\begin{equation}
\tuple{ G^i_{\tuple{p, \mathbf X_p}}} = \paren {\begin{array}{c}
v^\mu \\ 
- \map {\Gamma^\mu_{~\alpha \beta}}{x^\rho} v^\alpha v^\beta
\end{array}
}.
\end{equation}
Its connection to geodesics is best understood in the following way.
 Suppose we want to find a curve in $TM$ whose projection in $M$ is 
 a geodesic. This can be stated in the language of the geodesic spray 
 as the 1st order initial value problem:
\begin{equation}
\begin{cases}
\map {\tuple{\gamma^\mu, \dot \gamma^\mu}} {\lambda_0} &= \tuple {x^\mu_0, v^\mu_0}\\
\dfrac{d}{d \lambda} \tuple{\gamma^\mu, \dot \gamma^\mu} &= G^i_{\tuple{\gamma, \dot \gamma}}.
\end{cases}
\end{equation} 
In other words, all the integral curves of $\mathbf G$ 
are lifts of geodesics in $TM$.

The geodesic spray has a few other special properties. In local basis we have
\begin{equation}
\label{eq:gs}
\mathbf G_{\tuple {p, \mathbf X_p}} = v^\mu \dfrac{\partial}{\partial x^\mu} 
- \Gamma^\mu_{~\alpha\beta}v^\alpha v^\beta \dfrac{\partial}{\partial v^\mu} 
= v^\mu \mathbf e_{H\mu}
\end{equation}
Hence, $\mathbf G_{\tuple {p, \mathbf X_p}}$ belongs to the horizontal 
subspace $H_{\tuple {p, \mathbf X_p}}$. Furthermore, $d \pi_M \mathbf
 G_{\tuple{p, \mathbf X_p}} = \mathbf X_p$. Comparison with \eqref{eq:gf} 
 reveals that $\mathbf G$ generates the geodesic flow:

\begin{equation}
    \mathbf G_{\tuple {p, \mathbf X_p}} = \valat {\frac{d \map {\phi_\lambda} {\tuple{p, \mathbf X_p}}}{d \lambda}} {\lambda \mathop = \lambda_0}
\end{equation}


Geodesic spray is also involved in the formulation of various conservation laws on the tangent bundle \cite{sommers1973,prince1984a,prince1984b}. For example, consider the mapping $\map g {v, v} : TM \ni \tuple {x^\mu, v^\mu} \to \map {g_{\mu\nu}} {x^\rho} v^\mu v^\nu \in \R$, i.e. the norm of $v^\mu$. Then taking Lie derivative of $\map g {v, v}$ along $\mathbf G$ yields
\begin{equation}
\begin{split}
    \map {\calL_{\mathbf G}} {\map g {v, v}}  &= v^\mu g_{\alpha\beta,\mu}v^\alpha v^\beta - 2 \map {\Gamma^\mu_{~\alpha\beta}} x v^\alpha v^\beta g_{\mu\rho}v^\rho \\
    &= g_{\alpha\beta,\mu} v^\mu v^\alpha v^\beta - \paren{g_{\alpha \rho, \beta} + g_{\rho \beta, \alpha} - g_{\alpha \beta, \rho}}v^\rho v^\alpha v^\beta\\
    &= 0.
\end{split}
\end{equation}

\begin{figure}
    \includegraphics[scale=0.4]{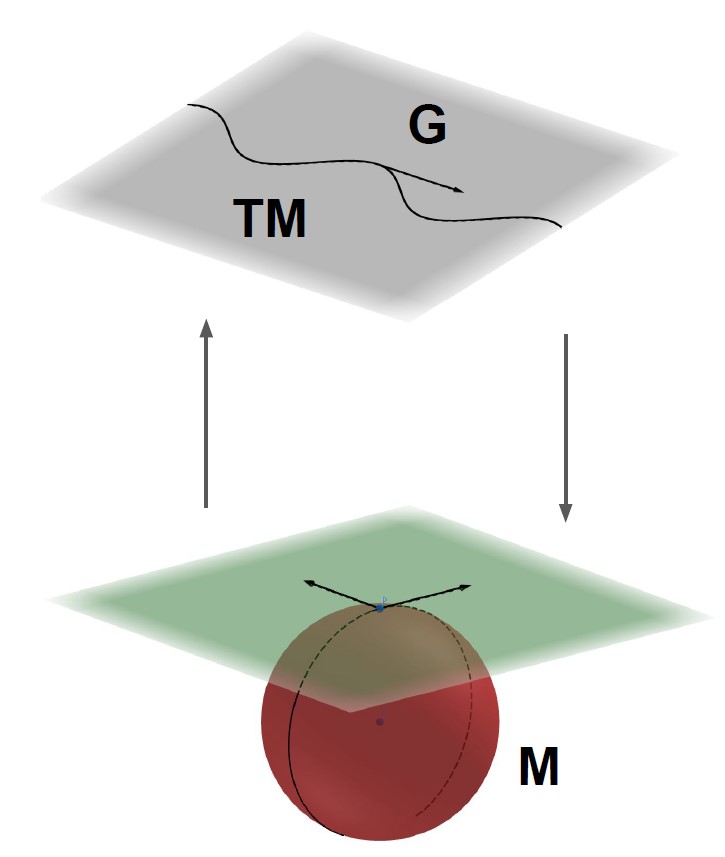}
    \centering
    \caption{Addendum: an illustration of the lift and projection of geodesic motion on the base manifold $\bf M$ and its tangent bundle $\bf {TM}$.
    The sphere represents the base manifold, the great circle - its geodesic, and the plane tangent to the sphere at a point of the great circle - the tangent space at a point of $\bf M$ containing
    the tangent vector of the geodesic as well as all the other tangent vectors at that point. The lift of this curve to the tangent bundle
    is given by the integral curve of the geodesic spray $\bf G$.}
\end{figure}

Another important structure is the sympletic form $\omega$ on $TM$, whose expansion in local cobasis reads 
\begin{equation}
\label{eq:smp}
\begin{split}
    \omega &= d v_\mu \wedge d x^\mu \\
    &= \map d {\map {g_{\mu\nu}} xv^\nu} \wedge dx^\mu\\
    &= v^\nu g_{\mu\nu,\rho} dx^\rho \wedge dx^\mu + g_{\mu\nu} dv^\nu \wedge dx^\mu.
\end{split}
\end{equation}
Again we are interested in its flow along $\mathbf G$. Lie derivative acts on differential forms by
\begin{equation}
\map {\calL_{\mathbf G}} \omega = \map d {\mathbf G \iprod \omega} + \mathbf G \iprod d \omega.
\end{equation}
The second term vanishes due to Eq. \eqref{eq:smp}. As for the first term, we have
\begin{equation}
\begin{split}
    \mathbf G \iprod \omega &= v^\nu g_{\mu\nu, \rho} v^\rho dx^\mu -v^\nu g_{\mu\nu, \rho} dx^\rho v^\mu - g_{\mu\nu} dv^\nu v^\mu - \Gamma^\nu_{~\alpha \beta} v^\alpha v^\beta g_{\mu\nu} dx^\mu\\
    &= \map d {-\frac 1 2 g_{\mu\nu} v^\alpha v^\beta}
\end{split}
\end{equation}
Altogether, $\map {{\calL}_{\mathbf G}} \omega = 0$, and the symplectic form is conserved.

 In semi-Riemannian geometry $\map g {v, v}$ and $\omega$ are always preserved along the geodesic spray. However, it may happen that the manifold has additional symmetries generating additional conservation laws. Consider a mapping $\xi : TM \ni \tuple {x^\mu, v^\mu} \to v^\mu \map {\xi_\mu}{x^\rho} \in \R$ where $\xi^\mu$ is so far an unspecified vector field on $M$. We are interested in finding such $\xi^\mu$ for which the Lie derivative of $v^\mu \xi_\mu$ along $\mathbf G$ vanishes. By a straightforward calculation we find that
\begin{equation}
\label{eq:kil}
    \map {\calL_{\mathbf G}} {v^\mu \xi_\mu} = v^\alpha v^\beta \xi_{\paren {\alpha; \beta}}.
\end{equation}
This will vanish for arbitrary $v^\mu$ iff $\xi_{\paren {\alpha; \beta}} = 0$. In other words, Eq. \eqref{eq:kil} holds for all geodesics iff $\xi^\mu$ is a Killing vector field on $M$. A similar argument for higher-order contractions results in equations for Killing tensor fields. Further extensions of this argument are possible by confining oneself to a subbundle. In the case of null geodesics, i.e. the subbundle $S^0 TM$, one can weaken the requirement by demanding vanishing Lie dragging only along lifts of null geodesics:
\begin{equation}
    \map {\calL_{\mathbf G}} {v^\mu \xi_\mu} = \map \phi {x^\rho} g_{\mu\nu}v^\mu v^\nu,
\end{equation}
where $\phi : M \to \R$ is an arbitrary smooth function.
Then instead of Killing equations, one obtains conformal Killing equations.
\section{Lie dragging of vector along $\mathbf G$}
\label{sec:liedr}

With all ingredients in place, we can 
now define the BGOs using the geodesic
spray, the geodesic flow it generates, and its deformation tensor $W^i_{~j}$. 
We begin by showing an equivalence between the Lie dragging of a
vector in $T_{\tuple {p, \mathbf X_p}}TM$ along the geodesic 
spray and the GDE. By doing so we will automatically reduce the GDE 
to the first order linear ODE. This in turn will allow us to express
the solution with the help of the resolvent operator, that is, 
a linear operator 
encapsulating all possible propagation effects of families of particles
between $2$ regions. In the last step, we will limit ourselves to the 
problem of propagation of light and demonstrate the connection between the geodesic flow and the BGOs in agreement with 
\cite{grasso2019}.

Let $\map \gamma \lambda$ be a geodesic with initial conditions
$\map \gamma {\lambda_0} = p$ and $\map {\dot \gamma} {\lambda_0} = \mathbf X_p$, and 
let $\map {\phi_\lambda} {p, \mathbf X_p}$ be its image under the geodesic flow, i.e. the lift of the geodesic. 
Let $\mathbf Y \in T_{\tuple {p, \mathbf X_p}} TM$  be a vector that
we drag along the integral curve of $\mathbf G$ to $T_{\tuple{\map \gamma
\lambda, \map {\dot \gamma} \lambda}} TM$. We would like to 
describe the geometry of neigbouring geodesic, at the lowest,
linear order in terms of the perturbation vector $\mathbf Y$. In the language of 
differential geometry is equivalent to requiring that $\mathbf Y$ is 
Lie dragged in the direction of $\mathbf G$. Then setting $\map 
{\calL_{\mathbf G}} {\mathbf Y} = 0$ yields
\begin{equation}
\label{eq:ldgs}
    \dfrac{d Y^i}{d \lambda} = G^i_{~,j} Y^j.
\end{equation}
Since $\mathbf G$ contains Christoffel symbols, one may expect that 
after differentiation, covariant structures like the Riemann
curvature tensor should emerge. Indeed, from Eqs. \eqref{eq:hvid}, \eqref{eq:gs}, the relation 
$\dfrac{d}{d \lambda} = G^i \partial_i$ and careful decomposition 
of the equation to the
horizontal and vertical subspace it follows that 
Eq. \eqref{eq:ldgs} is equivalent to the following ODE solved 
on the base manifold along the geodesic $\gamma$:
\begin{equation}
\label{eq:prop}
    \paren{\begin{array}{c}
         \nabla_{\dot\gamma} Y^\mu_H  \\
         \nabla_{\dot\gamma} Y^\nu_V
    \end{array}} =
    \paren{\begin{array}{cc}
       0  & \delta^\mu_{~\beta} \\
       R \UDDD \nu {\dot \gamma}{\dot \gamma} \alpha  & 0 
    \end{array}}
    \paren{\begin{array}{c}
         Y_H^\alpha \\
         Y_V^\beta 
    \end{array}}
\end{equation}
where $R^\nu_{~\dot \gamma \dot \gamma \mu} = R^\nu_{~\alpha\beta\mu}{\dot \gamma}^\alpha {\dot \gamma}^\beta$. This equation is linear both in $Y^\mu_H$ and $Y_V^\mu$ and defines a system of linear ODEs whose solution is expressible in terms of a resolvent operator. That is, the solution can be expressed as $\map {Y^i} \lambda = \paren {W_\lambda}\UD i j \map {Y^j} {\lambda_0}$.
Motivated by this, we introduce a map 
\begin{equation}
    W_\lambda : T_{\tuple{p, \mathbf X_p}} TM \ni \mathbf Y \to \map {W_\lambda}{\mathbf Y} \in T_{\tuple{\map \gamma \lambda, \map {\dot \gamma} \lambda}}TM
\end{equation}
satisfying the system of ODEs
\begin{equation}
\dfrac{d W^i_{~k}}{d \lambda} = G^i_{~,j} W^j_{~k} \label{eq:dWG}
\end{equation}
with initial data
\begin{equation}
\label{eq:wid}
\map {W^i_{~k}} {\lambda_0}= \delta^i_{~k} = \paren{\begin{array}{cc}
         \delta^\mu_{~\nu} & 0  \\
         0 & \delta^\mu_{~\nu}
    \end{array}}.
\end{equation}
On the other hand, we know that the tangent space of the tangent bundle is isomorphic to the direct sum of two tangent spaces:
\begin{equation}
T_{\tuple {p, \mathbf X_p}} TM \cong T_p M \oplus T_p M.
\end{equation} 
Together with the decomposition of vectors in horizontal and vertical parts, this suggests that $W_\lambda$ has an inherent 
$2 \times 2$ block structure. In the papers to be presented later, this is used extensively, but the notation is slightly 
different. We associate the horizontal part of $\mathbf Y$ with the perturbation of the position, and the vertical part -- 
with the perturbation of the tangent vector. Therefore, the subscripts for horizontal and vertical parts are replaced 
with $X$ and $L$ respectively, in line with the notation used in \cite{grasso2019,korzynski2022,serbenta2022}. 
Also, components $v^\mu$, and thus also the tangent vector to the geodesic, will now be denoted by $l^\mu$. 
Finally, we decompose $W_\lambda$ into $4$ operators that we label according to their connection with the initial and 
final data. 

\begin{figure}
    \includegraphics[scale=0.4]{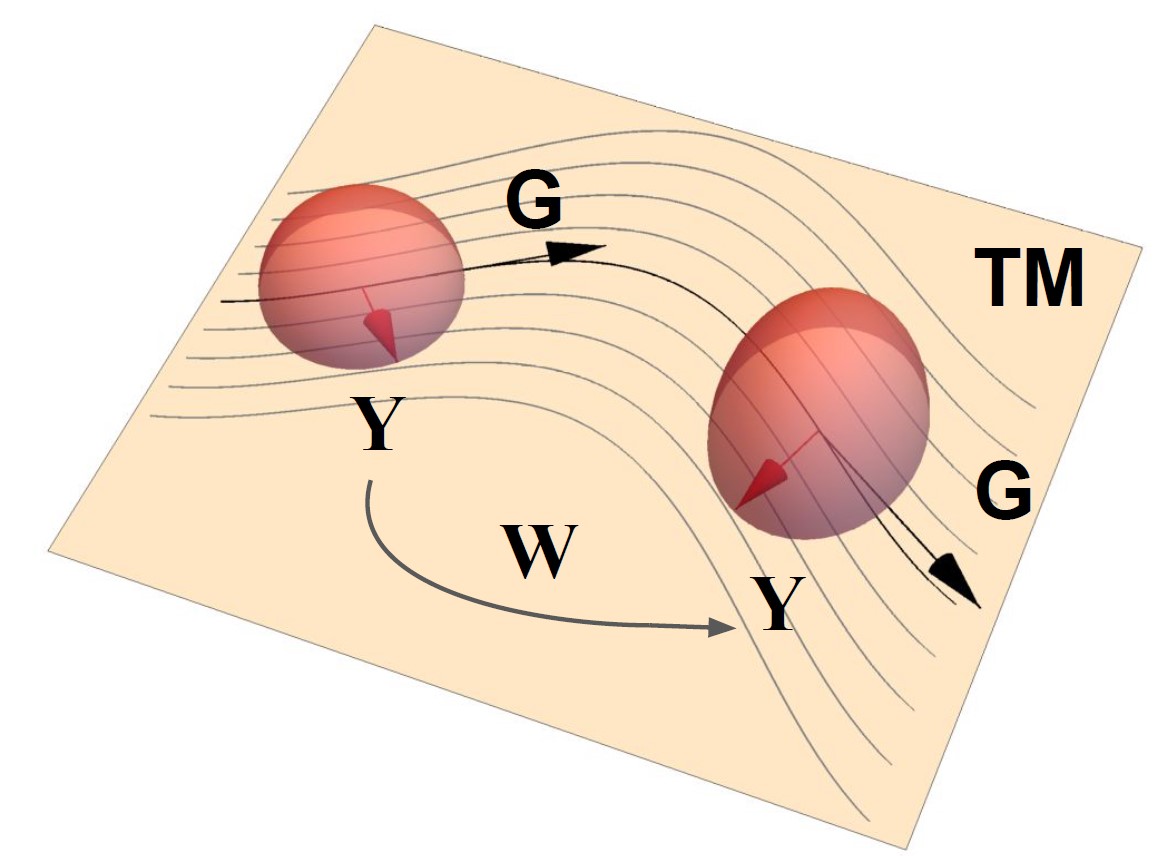}
    \centering
    \caption{Addendum: an illustration of the deformation of the geodesic flow generated by the geodesic spray $\bf G$.
    $\bf Y$ is an infinitesimal vector connecting two integral curves of geodesic sprays, and the span of all such
    $\bf Y$s define an infinitesimal domain. As we follow the reference integral curve, the deformation tensor $\bf W$
    varies as well. This determines the evolution of all $\bf Y$s and, accordingly, the shape of the infinitesimal domain.}
\end{figure}

Then the action of $W_\lambda$ can be expressed as 
\begin{equation}
\label{eq:res}
    \paren{\begin{array}{c}
         Y_X^\mu  \\
         Y_L^\mu 
    \end{array}} =
    \paren{\begin{array}{cc}
        W_{XX} {}\UD \mu\nu & W_{XL} {} \UD \mu \nu \\
        W_{LX} {} \UD \mu \nu & W_{LL} {}\UD \mu \nu
    \end{array}}
    \paren{\begin{array}{c}
         \map {Y_X^\nu} {\lambda_0}  \\
         \map {Y_L^\nu} {\lambda_0} 
    \end{array}}.
\end{equation}
Since both the vertical and the horizontal subspaces are isomorphic to the appropriate 
tangent space, the operators $W_{XX}$, $W_{XL}$, $W_{LX}$ and $W_{LL}$ can be identified with bitensors
acting between the tangent spaces at two different points:
\begin{equation}
    W_{**} \colon T_{\gamma(\lambda_0)} M \to T_{\gamma(\lambda)} M.
\end{equation}
The Eq. \eqref{eq:prop} can be rewritten as an equation for the covariant derivative of the bitensors
with respect to the point $\gamma(\lambda)$:
\begin{equation}
\label{eq:weq}
 \paren{\begin{array}{cc}
        \nabla_l W_{XX} {}\UD \mu\nu & \nabla_l W_{XL} {} \UD \mu \nu \\
        \nabla_l W_{LX} {} \UD \mu \nu & \nabla_l W_{LL} {}\UD \mu \nu
    \end{array}} = \underbrace{ \paren{\begin{array}{cc}
       0  & \delta^\mu_{~\beta} \\
       R \UDDD \mu ll \alpha  & 0 
    \end{array}}}_{S} \paren{\begin{array}{cc}
        W_{XX} {}\UD \alpha\nu & W_{XL} {} \UD \alpha \nu \\
        W_{LX} {} \UD \beta \nu & W_{LL} {}\UD \beta \nu
    \end{array}}
\end{equation}
Here we denote by $S$ a block matrix made of the curvature tensor and the unit matrix. It will later reappear in Eq. \eqref{eq:weqt}. From the block structure we also see that in principle the propagation of variations can be studied independently by appropriately setting the initial conditions, i.e. $\WXX$ and $\WLX$ depend only on initial position variations while $\WXL$ and $\WLL$ -- on initial direction variations.

We note here that Eq. \eqref{eq:weq} takes a particularly simple form when expressed in a parallel-transported tetrad along
the geodesic $\gamma$. The covariant derivatives become ordinary derivatives and we obtain simply a matrix
equation
\begin{equation}
\frac{d}{d\lambda} \paren{{\begin{array}{cc}
    W_{XX} {}\UD \mu\nu & W_{XL} {} \UD \mu \nu \\
     W_{LX} {} \UD \mu \nu & W_{LL} {}\UD \mu \nu
\end{array}}} = S \,\paren{\begin{array}{cc}
    W_{XX} {}\UD \alpha\nu & W_{XL} {} \UD \alpha \nu \\
    W_{LX} {} \UD \beta \nu & W_{LL} {}\UD \beta \nu 
\end{array}}. \label{eq:pptetrad}
\end{equation}
\section{Symplectic structure and BGOs}
\label{sec:symp}

We have previously shown that the geodesic spray $\mathbf G$ preserves the symplectic form $\omega$. With the help of the horizontal-vertical splitting we can now show the symplectic properties of BGOs.

Firstly, let us look at the action of $\omega$ on two arbitrary vectors $\mathbf Y, \mathbf Z \in T_{\tuple {p, \mathbf X_p}} TM$ with components $Y^i$ and $Z^i$. We know that on a semi-Riemannian manifold one can decompose vectors uniquely into their horizontal and vertical parts:
\begin{equation}
  \begin{split}
    \mathbf Y &= Y_H^\mu \mathbf e_{H\mu} + Y_V^\mu \mathbf e_{V\mu}\\
    \mathbf Z &= Z_H^\mu \mathbf e_{H\mu} + Z_V^\mu \mathbf e_{V\mu}.
  \end{split}
\end{equation}
By a straighforward calculation one can show that
\begin{equation}
\begin{cases}
\map \omega {\mathbf e_{V\mu}, \mathbf e_{V\nu}} &= ~~\map \omega {\mathbf e_{H\mu}, \mathbf e_{H_{\nu}}} = 0\\
\map \omega {\mathbf e_{V\mu}, \mathbf e_{H\nu}} &= - \map \omega {\mathbf e_{H\mu}, \mathbf e_{V\nu}} = g_{\mu\nu}
\end{cases}
\end{equation}
Then for any two vectors $\mathbf Y, \mathbf Z \in T_{\tuple {p, \mathbf X_p}} TM$ the symplectic form yields
\begin{equation}
  \begin{split}
    \map \omega {\mathbf Y, \mathbf Z} &= Y_V^\mu  Z_{H\mu} - Y_H^\mu Z_{V\mu}\\
    &= g_{\mu\nu} Y_V^\mu Z_H^\nu - g_{\mu\nu} Y_H^\mu Z_V^\nu\\
    &= \paren {Y_H^\mu, Y_V^\nu} \paren{\begin{matrix}
    0 & -g_{\mu\nu}\\
    g_{\mu\nu} & 0
    \end{matrix}
    }
     \paren{\begin{matrix}
     Z_H^\mu\\
     Z_V^\nu
     \end{matrix}
     }\\
     &= Y^i \Omega_{ij} Z^j
  \end{split}
\end{equation}
where $\Omega_{ij}$ plays the role of the skew-symmetric matrix. 
The explicit form of $\Omega_{ij}$ is important in deriving how the 
symplecticity of $W$ should be understood at the level of the block 
operators of $W$.

 We will now prove the simplecticity of $W$. Consider 
 the ODE for $W$ expressed in a parallel-transported tetrad, i.e. the matrix equation \eqref{eq:pptetrad}.  We take
 it together with its transpose:
\begin{equation}
\label{eq:weqt}
\begin{split}
\frac{d}{d\lambda} W &= S W\\
\frac{d}{d\lambda} \paren {W^T} &= \paren{W^T} \paren{S^T}
\end{split},
\end{equation}
where by transpose of $W$ and $S$ we mean
\begin{equation}
\paren{\begin{array}{cc}
       \WXX {}\UD \mu \nu  & \WXL {}\UD \mu \nu \\
        \WLX {}\UD \mu \nu & \WLL {}\UD \mu \nu
    \end{array}}^T = \paren{\begin{array}{cc}
       \WXX {}\DU \nu \mu  & \WLX {}\DU \nu \mu \\
        \WXL {}\DU \nu \mu & \WLL {}\DU \nu \mu
    \end{array}}
\end{equation}
and
\begin{equation}
\paren{\begin{array}{cc}
       0  & \delta {}\UD \mu \nu \\
        R \UDDD \mu l l \nu  & 0
    \end{array}}^T = \paren{\begin{array}{cc}
       0  & R_{\nu l l}^{~~~\mu} \\
        \delta \DU \nu \mu & 0
    \end{array}}.
\end{equation}
From Eq. \eqref{eq:weqt} it follows that
\begin{equation}
    \frac{d}{d\lambda} \paren {W^T \Omega W} = W^T \paren {S^T \Omega + \Omega S} W = 0.
\end{equation}
Hence, $W^T \Omega W$ must be constant along $l$. Finally we evaluate $W^T \Omega W$ at $\lambda = \lambda_0$
and take into account initial data from Eq. \eqref{eq:wid}. It follows that
\begin{equation}
\label{eq:symp}
W^T \Omega W = \Omega.
\end{equation}
Therefore, $W$ is symplectic. This also implies that not all components of $W$ are independendent. In terms of block operators Eq. \eqref{eq:symp} is equivalent to
\begin{equation}
\begin{cases}
\WLX {}_{\mu\rho} \WXX {}\UD \rho \nu - \WXX{}_{\mu\rho} \WLX {}\UD \rho \nu &= 0\\
\WLX{}_{\mu\rho} \WXL{}\UD \rho\nu - \WXX{}_{\mu\rho} \WLL{}\UD \rho\nu & = - g_{\mu\nu}\\
\WLL{}_{\mu\rho} \WXX{}\UD \rho\nu - \WXL{}_{\mu\rho} \WLX{}\UD \rho\nu & =  g_{\mu\nu}\\
\WLL{}_{\mu\rho} \WXL{}\UD \rho\nu - \WXL{}_{\mu\rho} \WLL{}\UD \rho\nu & = 0.
\end{cases}
\end{equation}

For completeness, we mention additional algebraic properties valid for smooth semi-Riemannian manifolds \cite{grasso2019}. From Eq. \eqref{eq:prop} one can see that for all $C, D \in \R$ the vector $\mathbf Y$ with the horizontal/vertical decomposition
\begin{equation}
\paren{\begin{array}{c}
         Y_X^\mu \\
         Y_L^\nu 
    \end{array}} = \paren{\begin{array}{c}
         \paren {C + D \lambda} l^\mu \\
          D l^\nu
    \end{array}}
\end{equation}
is a solution, where $l^\mu = \dot \gamma^\mu$ as explained in Sec.\ref{sec:liedr}. Assuming that $C$ and $D$ can be chosen independently, the substitution of $\mathbf Y$ into Eq. \eqref{eq:res} yields the following simultaneous set of constraints:
\begin{equation}
\begin{cases}
\map {\WXX{}\UD \mu\nu} \lambda\, \map {l^\mu}{\lambda_0} &= \map {l^\mu} \lambda \\
\map {\WXL{}\UD \mu\nu} \lambda\, \map {l^\mu}{\lambda_0} & = \paren {\lambda - \lambda_0} \map {l^\mu} \lambda \\
\map {\WLX{}\UD \mu\nu} \lambda\, \map {l^\mu}{\lambda_0} & = 0 \\
\map {\WLL{}\UD \mu\nu} \lambda\, \map {l^\mu}{\lambda_0} & = \map {l^\mu} \lambda
\end{cases}.
\end{equation}

In addition to this, any $\mathbf Y$ which solves Eq. \eqref{eq:prop} generates two conserved quantities:
\begin{equation}
\begin{split}
A &= \paren {Y^\mu_X - \lambda Y^\mu_L}l_\mu \\
B &= Y^\mu_L l_\mu.
\end{split}
\end{equation}
This can be checked by taking covariant derivatives of $A$ and $B$ along $l$ and using Eq. \eqref{eq:prop} to express $\nabla_l Y_X^\mu$ and $\nabla_l Y_L^\mu$ in terms of $Y_X^\mu$ and $Y_L^\mu$.
Evaluating them at both endpoints, applying Eq. \eqref{eq:res} and requiring the equalities to hold for arbitrary initial $\mathbf Y$ we get

\begin{equation}
\begin{cases}
\map {l^\nu} \lambda\, \map{\WXX {}\UD \mu \nu} \lambda & = \map {l_{\nu }}{\lambda_0}\\
\map {l^\nu} \lambda\, \map{\WXL {}\UD \mu \nu} \lambda & = \paren {\lambda - \lambda_0} \map {l_{\nu }}{\lambda_0}\\
\map {l^\nu} \lambda\, \map {\WLX {}\UD \mu \nu} \lambda & = 0\\
\map {l^\nu} \lambda\, \map{\WLL {}\UD \mu \nu} \lambda & = \map {l_{\nu }}{\lambda_0}.
\end{cases}
\end{equation}

\chapter{\emph{Bilocal geodesic operators in static spherically-symmetric  spacetimes}}\label{ssss}
In the first paper we present an exact solution of the GDE in terms of 
BGOs along any null geodesic in static, spherically symmetric spacetimes 
and its application to the  of the behaviour of the distance measures in the 
Schwarzschild spacetime.

The solution is derived in two different ways. The first approach requires
 a general solution of the GE, either in exact form or in terms of 
 implicit relations. Then one has to vary it with respect to the 
 components of the initial position and direction. After a few 
 adjustments, which make the expressions covariant, the solution 
 is recovered. The second approach makes direct use of 
 the symmetries of the spacetime. Namely, Killing vectors 
 generate the first integrals of the GDE, which effectively 
 reduces the order and the number of ODEs we need to solve. 
 In the end, the solution depends on various conserved quantities, 
 initial conditions and may possibly also involve nontrivial integrals 
 of functions of metric coefficients. Thus derived solutions, 
 expressed in the coordinate tetrad, are valid for timelike, spacelike
  and null geodesics whenever the chosen local coordinates are valid.

Still, the physical interpretation of the solution 
requires some care due to chosen coordinate system. 
The situation can be improved by rewriting the result 
in a different tetrad with a clear physical interpretation.
 In this paper we choose the parallel transported semi-null tetrad 
 (SNT). We first 
 construct it at the point of observation and then solve the
  parallel propagation equation while simultaneously taking 
  into account the constraints of the SNT. Then the solution 
  of the GDE is projected onto this SNT, in which the solution
   takes a much simpler form.

Finally, we use the results to study the angular diameter
and the parallax distances in the spacetime of Schwarzschild 
black hole. The behaviour is analyzed numerically while the 
most apparent properties are explained qualitatively by splitting 
the problem into three different regions: the initial region close
 to the observer, the intermediate region around the black hole, 
 and the faraway region with the source positioned at increasingly
  larger distances from the black hole.

Although the paper discusses only static spherically 
symmetric spacetimes, the principles behind the methods 
of solution can be readily generalized to any sufficiently
 symmetric spacetime. In other words, if a spacetime possesses
  an adequate number of Killing vectors and tensors, both methods
   will produce a solution, at least in the coordinate tetrad.
    Furthermore, the projection of the results on the SNT requires
     the parallel transport of the SNT. In this 
     paper, the result was achieved by assuming an 
     ansatz motivated by the assumed geometry of the 
     spacetime and then fulfilling the SNT constraints.
      When the form of the ansatz is not self-evident, one 
    can still avoid solving the parallel transport 
    equations if the spacetime admits a Killing-Yano tensor. 
    However, this is not always guaranteed, not even in the presence 
    of a Killing tensor. In that case one is left with a system of 
    linear ODEs.

\section*{Author's contribution}

The results were derived in collaboration with my supervisor prof. M. Korzyński whose contribution includes:

\begin{itemize}
\item the suggestion of the 1st algorithm;

\item the idea of the investigation of distance measures in the black hole spacetime

\item some help with the refinement of sections 3.3, 4.1 and 4.4-4.6.
\end{itemize}

Meanwhile, my contribution includes:
\begin{itemize}
\item the suggestion and the implementation of the 2nd method;

\item all the derivations;

\item numerical investigation and the plots.
\end{itemize}
The draft was written and published jointly by me and M. Korzyński.

\section*{Addendum}

One of the referees correctly pointed out that fixing $\map C r = r^2$ in Eq.(23)
actually reduces the generality. Nariai solution of Einstein equations \cite{nariai1950,nariai1950rep,nariai1951new,nariai1951rep} and Bertotti-Robertson
solution of Einstein-Maxwell equations \cite{bert1959,rob1959} do have the property that $\map C r = const$.
Luckily, this does not cause any radical changes in the derivation.

Another mistake concerns terminology. The name "surface-forming" has to be replaced with
"hypersurface-orthogonal". Two vector fields, $u$ and $v$ are
called surface-forming if their integral lines form congruence of surfaces. Then they satisfy the equation
\begin{equation*}
    [u, v] = A u + B v \qquad \textrm{or} \qquad \epsilon_{\alpha\beta\mu\nu} u^\beta v^\mu \mathcal{L}_u v^\nu = 0
\end{equation*}
where $A$ and $B$ are scalar coefficients. On the other hand, hypersurface-orthogonal vector field
$v$ satisfies:
\begin{equation*}
    v_\alpha = \lambda f_{,\alpha} \qquad \textrm{or} \qquad \epsilon^{\alpha\beta\mu\nu} v_{\beta;\mu} v_\nu = 0
\end{equation*}
where $\lambda$ is a scalar coefficient and $f$ is a scalar function \cite{stephani2009}.

 \includepdf[scale=1.2, pages=-]{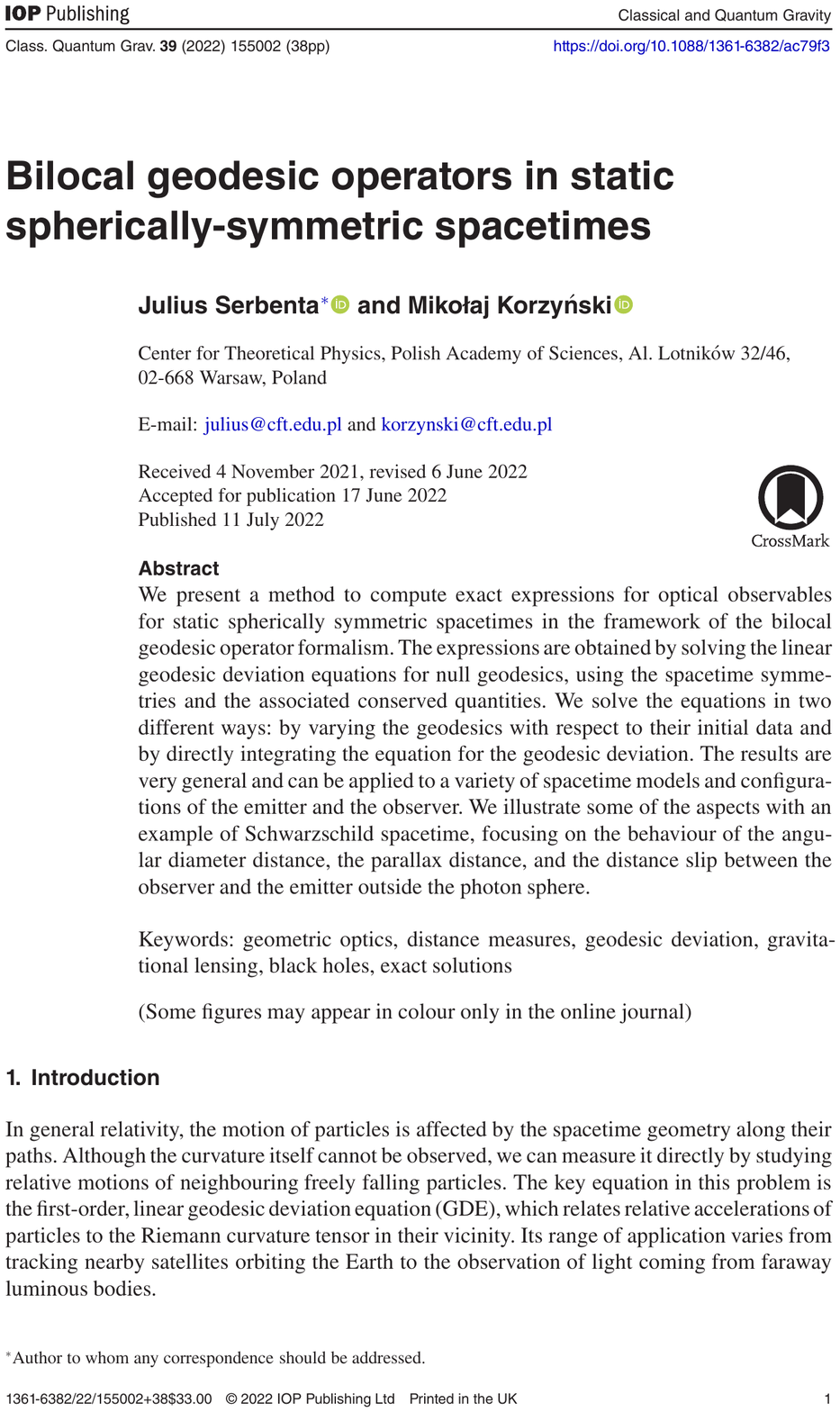}
\chapter{\emph{Testing the null energy condition with precise distance  measurements}}\label{testn}
In this chapter, we present a very general and unexpected result 
about the distance measurements based on light propagation in curved spacetimes. It states that if
general relativity holds, the matter satisfies the NEC
 and the light travels along null geodesics, then for a source
  of light placed anywhere between the observer and the first focal point the parallax distance measured 
   by the observer cannot get smaller than the angular
    diameter distance. The result is completely independent
     of any spacetime symmetries or motions of the source or the observer. 
    The result rests upon the projected GDE and the light ray 
    bundle formalism, while the techniques used are very similar
 to those found in standard focusing theorems.

In this paper we study the 
inequality with three different parallax distance definitions:
 the determinant-averaged, the trace-averaged,
  and the directional parallax distances. The result holds equally 
  for the first two definitions, although the proof of the second 
  one is more technically involved than of the first one. On 
  the other hand, the directional 
  parallax distance is shown not to satisfy the inequality if the 
  ray encounters strong tidal forces. The paper is concluded
   with a list of rough estimates of the positions of focal points
    and the distance slip throughout and beyond the Milky Way. For 
    this we assumed a certain continuous matter distribution model. 
    We conclude that direction averaging is crucial unless Weyl 
    contribution is on average negligible and uncorrelated with the direction of the baseline of observation.

\section*{Author's contribution}

The results were derived in collaboration with my supervisor prof. M. Korzyński. My contribution includes:

\begin{itemize}
\item making the connection between the BGO and the lightray bundle formalisms, especially the fact that $\mu$ is related to the lightray bundle with initially parallel rays;

\item estimation of $\mu$ in Sec. IV;

\item development of Taylor series expansion for BGOs and distance measures used in Sec. III;
\end{itemize}

Meanwhile, the contribution of prof. M. Korzyński includes:
\begin{itemize}
\item the idea of the paper;
\item the complete proof for the inequality of trace-averaged parallax distance and the counterexample in the case of strong Weyl curvature in Sec. III.
\end{itemize}
The draft was written and published jointly by me and M. Korzyński.

 \includepdf[pages=-]{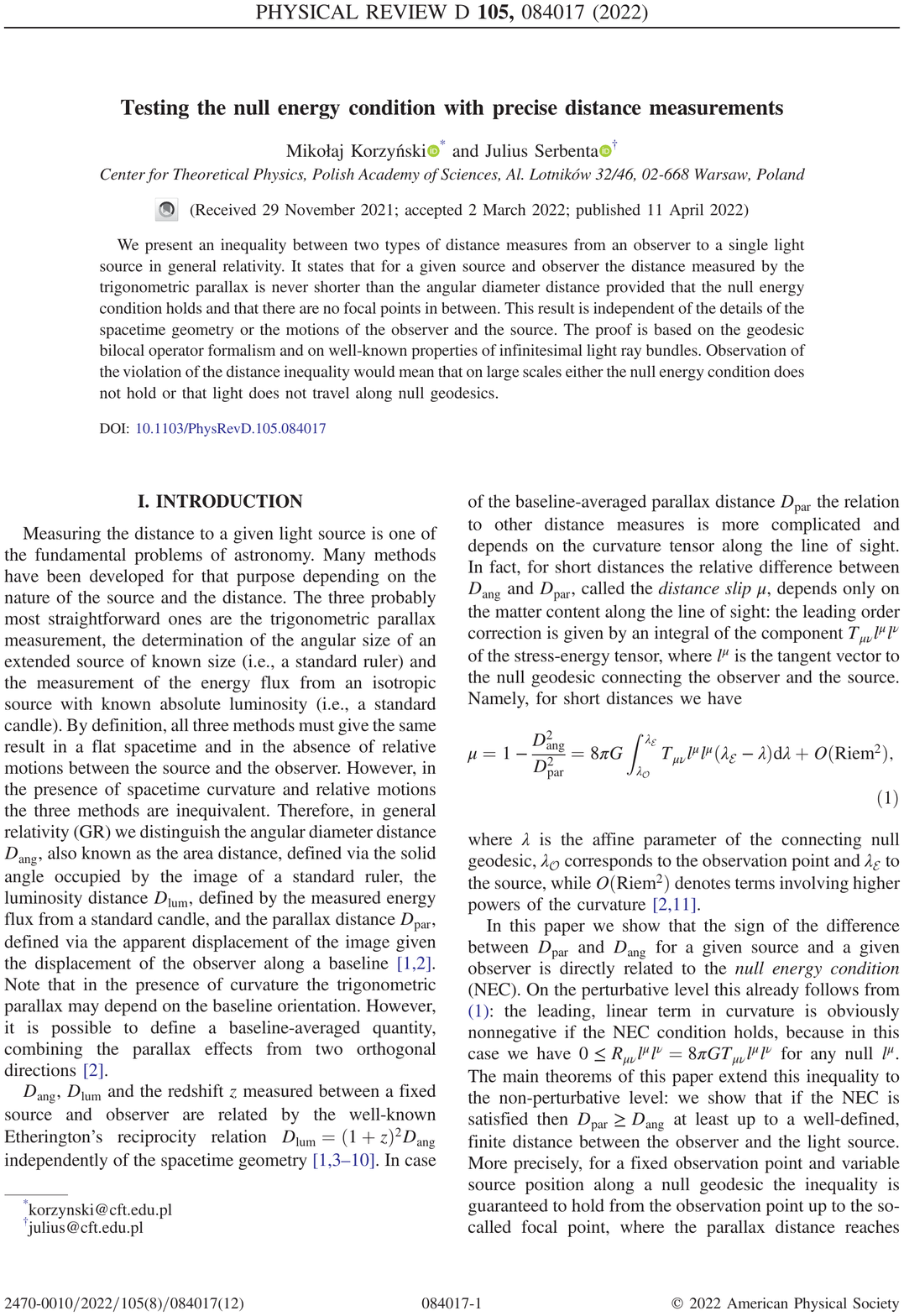}

\chapter{Addendum: observational prospects of distance slip and drift effects}

The BGO formalism extends the previous approach to the geometrical optics in
General Relativity by considering the entire dimensionality of the problem together
with the contributions coming from the observer's and emitter's peculiar
motions. Therefore, it is capable of describing more optical effects. However,
not in all cases the BGO formalism can be applied straightforwardly. 
We can consider the displacements of both the observer and the source, and the displacement 
may also be timelike, corresponding to the observer's time flow. The last point is crucial 
when discussing the drifts. The topic of drifts was considered in other papers of our group 
\cite{grasso2019,korzynski2018}, which did not make it to this PhD thesis.

\section{Redshift drift}

The redshift drift is a rather old prediction of the FLRW model \cite{sandage1962,loeb1998}. 
Although the very first investigation considered the effect to be non-measurable,
the recent work suggests that some of the new observatories will be capable of
achieving this. In reality, the idealized cosmological signal will be contaminated
with the kinematical contribution, possibly coming from the bulk 
flow, and the deviation from homogeneous mass distribution. $\mathrm{\Lambda CDM}$ predicts 
$|\delta z| \approx 10^{-10}$ (or $|\delta v| \approx 1 \mathrm{cm/s}$, or $\delta f = 0.1 \mathrm{Hz}$) 
per ~10 years. In practice, we can expect the following:

\begin{itemize}
    \item (from Lyman-$\alpha$): ELT with its 40m diameter dish could measure the effect
    from Lyman-$\alpha$ forest of radiation emitted by quasars at z = 4 in 10-20
    years \cite{uzan2008time, pasquini2005}

    \item (from hydrogen lines): SKA should measure redshifts with the accuracy of $10^{-6}$ by observing
    21 cm Hydrogen lines (HI). Measurement of redshift drift and cosmic jerk will take 12 and 50 years respectively

    \item (from drift of gravitational lens system parameters): for a multiply-lensed quasar at $z = 1.4$ with a lensing cluster at $z = 0.36$ we
    get angular drift of 0.1 nas/year and time delay drift of 1 ms/year \cite{piattella2017,covone2022} - not
    measurable
    \item (from differences of redshifts of gravitationally lensed images): for a galaxy cluster with velocity
    dispersion $\sigma_v = 1500 \mathrm{km/s}$, eccentricity $\epsilon  = 0.6$, redshifts of the lens and the source $z_l = 0.5$, $z_s = 4$ respectively, we can
    reach $\delta z \approx 10^{-8}$ \cite{wang2022,helbig2023}
\end{itemize}

\section{Parallax and position drift}

There is no single well-accepted definition of the parallax both in theoretical
and experimental contexts. In this thesis by parallax we mean the geometrical
parallax which is a momentary observation. It is related to the difference of
the apparent position on the sky of the source between observers displaced in
the transverse direction. On the other hand, the parallax of the moving source
which is measured over a period of time by a moving observer, we call the
position drift. Their distinction is important because in general they describe
different scenarios and measure different aspects of the spacetime geometry.

In actual observations we can identify at least two notions of the parallax.
The first type is the annual parallax. It is the effect seen by a single observer
on Earth observed in a year. Such parallax includes the proper motion and the
annual variation, whose magnitude is interpreted as the "momentary" parallax
effect. With current precision of observations and a relatively short baseline, it can
only be used within the Milky way (the fartherst measured parallax distance is 
$\approx 20 \mathrm{kpc}$ \cite{sanna2017}, while routine observations extend to less than half of that \cite{bailer2021}).

Another type of parallax is the secular parallax known as Kardashev parallax \cite{kardashev1986}.
It is the parallax due to motion of the Solar System wrt CMB frame, i.e. it is the secular
growth of the parallax signal. Every year the Solar system travels $\approx$ 80AU in roughly the same direction.
The parallax depends on the angle between the direction towards
the source and the direction of motion. 
However, the parallax signal (i.e. change of apparent position) is still very
small and mixed with secular aberration effects (Solar System motion around
the Milky Way center), peculiar motions, bulk flows, deviations from isotropic
and homogeneous expansion. Also, the baseline extends only in one direction.
This can potentially mask the true parallax due to curvature contribution \cite{korzynski2022},
but the problem can be averted by averaging over many sources. We note that the notion of the cosmic parallax varies 
between authors \cite{grasso2019}, and sometimes the cosmic parallax includes all of the aforementioned effects.

In perfect $\mathrm{\Lambda CDM}$ cosmology for matter comoving with the Hubble flow there is no position
drift \cite{krasinski2023}. A non-zero effect can arise in case of inhomogeneities and peculiar
velocities and accelerations \cite{marcori2018}. However, up to this point only $5 \mathrm{\mu as / yr}$ aberration
drift part of the position drift has been measured \cite{titov2011, xu2012, klioner2021} by observing 
extragalactic radio and QSO-like sources. The dipole of the drift is pointing roughly at the center of MW,
although a slight vertical component has also been reported. The result is consistent with the acceleration of the Solar
System in the Milky Way gravitational field.

In practice, we can expect the following:

\begin{itemize}
    \item Secular parallax: at z = 0.01 (or comoving distance of 40 Mpc) 1-2 $\mathrm{\mu as / yr}$ are expected \cite{kardashev1986,marcori2018}; $1 \mathrm{\mu a s}/ 10 \mathrm{yrs}$ for sources at $z = 1$.
    \item Gaia: in 5 years with 500,000 quasars one can expect 10-200 $\mu as$ sensitivity; in 10 years with $10^6$ sources - $\approx 5 \mu as$ \cite{quartin2010, quercellini2009, quercellini2009p}
    \item Gaia: two missions with 10 years appart may test 1 Gpc void hypothesis through the cosmic parallax \cite{fontanini2009, quercellini2012}
\end{itemize}

\section{Distance slip}

Within the Milky Way we may try to estimate the distance slip $\mu$. In this
case, the difference between the $D_{ang}$ and $D_{par}$ is at most $\approx 10^{-4}$ \cite{korzynski2022}. Hence,
a lot of objects ($\approx 10^4 - 10^5$) would be needed, for which we could very precisely measure
the annual parallax and the angular diameter distance (or the
luminosity distance and the redshift). The distance slip would give some idea
about matter density along the line of sight in the Galaxy. But currently it
seems beyond what we can do.

The estimate can also be done for the distance slip on cosmological scales. For example, in an FLRW model,
$\mu(z = 1) = 0.22$ \cite{villa2020}.

\section{CMB drift and parallax}

We close this chapter by a quick note about the evolution of the CMB signal, 
which although has not been discussed in our work, but in principle can be studied as well. 
CMB represents the intersection of our past light cone with the last scattering
surface. As time passes, the vertex of the light cone moves which also affects
the intersection, while the light cone also tracks a longer development of the
evolution of the spacetime geometry. Hence, we can expect the following changes
of CMB features in time \cite{lange2007}:

\begin{itemize}
    \item The average temperature should reduce as $a^{-1}$
    \item The peaks and the minimum should shift to smaller scales, because the viewing angle is shrinking
    \item Late ISW effect will grow wrt the first peak
    \item Our motion affects the dipole.
\end{itemize}

But almost all estimates claim that these effects are unobservable (or observable in $10^3$ years). More precisely, we have the
folowing estimates \cite{zibin2007, moss2008, quercellini2012}:

\begin{itemize}

    \item CMB monopole shift: $1 \mathrm{\mu K}$ in 5000 years; eventually it may be dominated by de Sitter 
    thermal noise, but this will be visible only in 1 Ty
    \item CMB dipole - Planck mission would take hundreds of years to observe it, while a 100x more sensitive mission would take $10$ years; 
    \item CMB quadrupole parallax in $\mathrm {\Lambda CDM}$ universe - a sensitivity of $10^{-4} \mathrm {\mu as}$ is needed
    \item observation of differences in higher multipoles would require 1000s of years for the signal to accumulate 

\end{itemize}

\chapter{Conclusion}

In this thesis I present the theory and applications of general relativistic transfer matrices for null geodesics known as the bilocal geodesic operators. The transfer matrix method is well known in studies of linear systems. In General Relativity the Jacobi propagators play the role of transfer matrices for deviations of null and timelike geodesics. However, the transfer matrix theory for the propagation of light as presented in the literature seems incomplete, as it cannot accommodate either redshift or position drift effects. The formalism presented here allows to consider all possible lowest-order nontrivial optical effects.

The bilocal geodesic operator formalism was originally presented as a resolvent operator for the geodesic deviation equation. In this thesis I present a different representation of the problem. Instead of working with the null geodesics on the base manifold, I consider their lifts to the tangent bundle. Then the dimensionality of the resolvent operator becomes twice as big, but the governing differential equations reduce from the 2nd order to the 1st order. Furthermore, the metric structure of the base manifold permits a covariant splitting of vector fields on the tangent bundle into two invariant parts -- the horizontal and the vertical subspace. It can be shown that each subspace is isomorphic to a copy of the tangent space to the base manifold. This implies that the resolvent operator can be decomposed into four distinct bilocal operators on the base manifold, known as the BGOs.

In the first paper we presented expressions for an analytic form of the BGOs for 4D static spherically symmetric spacetimes and applied them to the Schwarzschild spacetime. The solution was obtained in two ways. The first method requires differentiation of the solution to the geodesic equation with respect to its initial conditions. The result then needs to be updated for the final result to be fully covariant. The fix is based on the standard decomposition 
into the horizontal and vertical subspace of the tangent bundle. This decomposition is coordinate-invariant by construction. 

The second approach involves a direct integration of the GDE. Symmetries of the spacetime together with the symplecticity of the GDE give rise to the first integrals, or quadratures. The next step involves expressing the BGOs in a parallel-transported semi-null tetrad. It turns out that after the projections the expressions simplify dramatically, especially in the screen space. The results are then applied to the Schwarzschild spacetime to study the behaviour of the angular diameter and parallax distance measures as well as their associated distance slip. Depending on the impact parameter of the fiducial null geodesic we expect a formation of special points along the fiducial light ray, where distance measures either vanish or diverge.

In the second paper we prove an inequality between the parallax distance and angular diameter distance under rather general conditions. Namely, if we assume the validity of General Relativity, the null energy condition, and the propagation of light in vacuum, then it can be shown that for any given source of light the parallax distance defined by arithmetic or geometric mean cannot grow larger than the angular diameter distance. Averaging of the parallax effect over both linearly independent baselines is crucial in order to avoid contributions of the Weyl tensor at linear order, which break the inequality for certain directions of the baseline. The result is blind to the symmetries of spacetime, while the proof shares some of its methods with the well-known focusing theorems. However, experimental confirmation of this inequality seems to be too difficult. The measurement in the Milky Way requires precise measurements of distances to $10^5$ sources on the opposite side of the disk together with good modeling of the matter distribution in the galaxy. It is also possible to consider the cosmological parallax due to the motion of the solar system with respect to the CMB. Again, the expected signal with current technologies is too low, but the secular growth of the effect makes it in principle measurable.

\section*{Future directions}

The topic can be extended in several different ways. Firstly, in comparison with the number of exact solutions of Einstein equations, there is only a handful of metrics with general exact solutions to the geodesic deviation equation (see appendix \ref{exs}). Therefore, it would be interesting to find the BGOs for other spacetimes with symmetries and classify them by various optical effects. 

Secondly, position \cite{krasinski2011,krasinski2011a,krasinski2012,krasinski2012a} and redshift \cite{yoo2008,yoo2011,quartin2010,uzan2008,gasperini2011} drift effects have been considered only for a few models of the Universe. At the moment there are several observatories under construction \cite{quercellini2012,kloeckner2015,cristiani2007}, which will enable the measurement of redshift drift effects for the first time. A better understanding of drift effects and conditions under which they vanish is welcome.

Thirdly, some of the BGOs so far have been considered only formally, i.e. as a part of the total resolvent operator. It is still not clear whether they or their combinations correspond to any observable effects. Moreover, it is known that the Jacobi map satisfies Etherington's duality relation which holds for all Lorentzian spacetimes with Levi-Civita connection \cite{uzun2020}. One may wonder if there are more duality relations expressible in terms of all the BGOs.

In practice, the spacetime geometry of the Universe cannot be expressed as an exact solution to Einstein's equations or their modifications. Instead, we have to assume a parametrized model and find the best fit with the least possible errors. Any statement about a well-fitted model has to be corroborated with an appropriate statistical analysis of the observational data \cite{peebles2020,durrer2020,schneider2006,martinez2009,verde2010,fleury2015}. Nonetheless, the lowest-order perturbation theory and its statistical analysis both in the context of cosmological perturbations and weak gravitational lensing provide an extremely good agreement between the theory and observations. An equivalent perturbative and statistical study of new optical effects, such as the drifts, and their connections to the standard observables could make it easier to bring the formalism to practice. The robustness of data interpretation can be further increased by reducing its dependence on the model assumption and enforcing covariant formulation of the formalism. Recent studies suggest that redshift drift cosmography admits a metric-independent cosmological data analysis \cite{heinesen2022p,koksbang2022r,heinesen2021r2,heinesen2021r1,heinesen2021m}. Extension of these ideas to other observables, such as the position drift, could be interesting.

The applicability of the formalism relies heavily on the measurability of new optical observables. The spatial resolution of individual light sources is possible only in the immediate vicinity of the Solar system and during strong lensing scenarios. On the other hand, the time of arrival of emitted particles can be measured with much higher precision. This serves as the motivation for the local surface of communication introduced in \cite{korzynski2021}. In this article, the BGOs are replaced by a different set of operators mapping position perturbations at both endpoints to their vector perturbations, but both sets of operators are equivalent in the sense that they involve the same functionals and variables. Even so, not all properties of the BGOs have been translated to the properties of the local surface communication. Particularly intriguing is the connection between the tangent bundle structure in the BGO formalism and the product manifold structure in \cite{korzynski2021}.

Finally, the BGO formalism only describes the linear geodesic deviation equation. The full geodesic deviation is much more involved due its nonlinearity in terms of curvature tensors \cite{clarkson2016,clarkson2016a}\footnote{Motivation for this can be given by taking a difference of integral curves of $\mathbf G_{\tuple{p, \mathbf X_p}}$ evaluated at positions differing by a vector $\mathbf Y_p$ and expanding the difference around the reference integral curve, which yields \\ $\dot Y^i = \sum_{k = 1}^\infty \frac 1 {k!} G^i_{~,j_1 \ldots j_k} Y^{j_1}\ldots Y^{j_k}$.} and path deviations as well as its non-locality. Still, the geometrical picture presented in Ch.\ref{mathin} could work as a starting point to generalize the geodesic deviation beyond the normal convex neighbourhood in a covariant manner \cite{vines2015,schattner1981,harte2012,dixon1979,dixon1974}

\bibstyle{abbrv}
\printbibliography

\appendix

\chapter{Derivation of GDE from geodesic flow}

In this section we give a more precise derivation of Eq. \eqref{eq:prop} from Eq. \eqref{eq:ldgs}. Recall that in the basis $\set {f_i} = \set{\mathbf f_{x,\alpha}, \mathbf f_{v,\beta}}$ (Eq. \eqref{eq:trb}) the geodesic spray reads

\begin{equation}
\paren {G^i} = \paren{\begin{array}{c}
v^\alpha \\
- \map{\Gamma^\beta_{~\mu \nu}} {x^\rho} v^\mu v^\nu 
\end{array}}.
\end{equation}
On the tangent bundle we have the induced coordinate system $\tuple {\chi^i} = \tuple {x^\kappa, v^\delta}$. We choose the ordering of the partial derivatives to match the ordering of the coordinates. Then the partial derivatives of $G^i$ can be listed as
\begin{equation}
\paren {G^i_{,j} } = \paren{\begin{array}{cc}
0 & \delta^\alpha_{~\delta} \\
- \map{\Gamma^\beta_{~\mu\nu,\kappa}}{x^\rho}v^\mu v^\nu & -2 \map {\Gamma^\beta_{~\mu \delta}}{x^\rho} v^\mu
\end{array}}.
\end{equation}
Now return to Eq.\eqref{eq:ldgs} and write out the ODEs for $Y^i$ expressed in the associated basis $\set {f_i}$:
\[\arraycolsep=1.4pt\def\arraystretch{2.2}
\paren{\begin{array}{c}
\dfrac{d}{d \lambda} Y^\alpha_x \\ \dfrac{d}{d \lambda} Y^\beta_v
\end{array}} = \paren{\begin{array}{c}
Y^\alpha_v \\ - \Gamma^\beta_{~\mu\nu,\kappa}v^\mu v^\nu Y^\kappa_x - 2 \Gamma^\beta_{~\mu\delta}v^\mu Y^\delta_v
\end{array}}.
\]


As it was explained in Sec. \ref{sec:geomtb}, the splitting can be made invariant with respect to coordinate transformations \eqref{eq:indtr} by the change of variables \eqref{eq:hvid}. After a bit of calculation and term rearranging the equations can be rewritten as
\[\arraycolsep=1.4pt\def\arraystretch{2.2}
\paren{\begin{array}{c}
\dfrac{d}{d \lambda} Y^\alpha_H +\Gamma^\alpha_{~\mu\nu} Y^\mu_H v^\nu \\ \dfrac{d}{d \lambda} Y^\beta_V + \Gamma^\beta_{~\mu\nu} Y^\mu_V v^\nu
\end{array}} = \paren{\begin{array}{c}
Y^\alpha_V \\ R \UDDD \beta \mu \nu \rho v^\mu v^\nu Y^\rho_H
\end{array}},
\]
where $R \UDDD \beta \mu \nu \rho$ is a collection of Christoffel symbols and their derivatives constituting the components Riemann tensor on the base manifold. Note that the terms on the left-hand side cannot be yet reduced to the covariant derivatives, because these expressions are components of a vector in $T_{\paren{p, \mathbf X_p}} TM$, and $TM$ has not been supplied either with a metric or a connection.

The final step of reduction is the utilization of the isomorphisms $\operatorname {iso}^H$ and $\operatorname {iso}^V$ \eqref{eq:iso}. This way we can make the following identifications:
\begin{equation}
\begin{split}
T_{\tuple {p, \mathbf X_p}} TM\ni Y^\alpha_{H}\, \mathbf e_{H \alpha} &\longrightarrow \tilde Y_{H}^\alpha\, \frac{\partial}{\partial x^\alpha} \in T_p M \\
\dfrac{d Y^\alpha_H}{d \lambda} + \Gamma^\alpha_{~\mu\nu}v^\mu Y^\nu_H &\longrightarrow \dfrac{d \tilde Y^\alpha_H}{d \lambda} + \Gamma^\alpha_{~\mu\nu} l^\mu \tilde Y^\nu_H \equiv \nabla_l \tilde Y^\alpha_H,
\end{split}
\end{equation}
where $\tilde Y_H$ is a vector field along the
geodesic $\gamma$, whose components 
are equal to those of $Y_H$, i.e.
$\tilde Y_H^\alpha = Y_H^\alpha$. 
A similar construction can be used for the vertical part. Therefore the isomorphisms allow us to regard the equations for horizontal and vertical parts as proper expressions on the spacetime $M$. 
$M$, on the other hand, is equipped with a metric tensor and a covariant derivative. As a result, we end up with a system of equations on the base manifold
\begin{equation}
\begin{cases}
\nabla_l \tilde Y_H^\alpha &= \tilde Y_V^\alpha\\
\nabla_l\tilde Y_V^\beta &= R \UDDD \beta \mu \nu \rho l^\mu l^\nu \tilde Y_H^\rho.
\end{cases}
\end{equation}
(recall that $v^\mu = \dot \gamma^\mu$, because
we are dealing with a lift of a geodesic, and we have introduced the notation $l^\mu = \dot \gamma^\mu$).
This system is equivalent to the GDE for $\tilde Y^\alpha_H$.
We can check that by substituting $\tilde Y_V$ in the second equation with the expression from the first one, obtaining  this way $\displaystyle \nabla_l \nabla_l \tilde Y_H^\alpha = R \UDDD \alpha \mu \nu \rho l^\mu l^\nu \tilde Y^\rho_H$.

\chapter{List of exact solutions to GDE}
\label{exs}

The geodesic deviation equation constitutes a set of coupled linear second order ordinary differential equations. Although the analysis of ODEs is simpler than that of partial differential equations, finding a compact solution to a system of ODEs is not an easy task \cite{Cariglia2018erv,dryuma1997}. However, most contemporary articles contain a very limited list of examples of exact solutions. In this appendix we provide a list of spacetimes for which the corresponding GDE has been integrated analytically.

\begin{itemize}

\item Constant curvature spacetimes: \cite{synge1934deviation,synge1960,peters1969covariant}

\item Spherically symmetric spacetimes (without deriving the full BGOs): \cite{nariai1962invariant}

\item Static spherically symmetric spacetimes: \cite{nariai1960some,dyer1977,fuchs1984,
fuchs1990,fuchs1990par,serbenta2022,
sokolowski2014local,sokolowski2014local2,sokolowski2019globally}

\item Schwarzschild spacetime: \cite{newman1959measurement,manasse1963fermi,dwivedi1972,shirokov1973one,peters1976,cooperstock1979local,bazanski1986,bazanski1989,mlodzianowski1989,kerner2001,philipp2015}

\item Newtonian gravity with oblate Earth: \cite {greenberg1974equation}

\item Reissner–Nordström spacetime: \cite{balakin2000,Gad2007GeodesicsAG,sokolowski2014local2}

\item Kerr spacetime: \cite{dolan1984solutions,kostyukovich1985,Cariglia2018erv}

\item FLRW spacetime: \cite{newman1959measurement,
neville1983compact,kasai1988tr,ellis1999,villa2020}

\item conformally flat spacetime \cite {peters1975}

\item plane symmetric cosmological models: \cite{singh1968,singh1973}

\item Bianchi cosmological models: \cite{roy1982,Roy1985Bianchi,roy1990bian}

\item pp-wave: \cite{balakin2000}

\item rotating disk spacetime: \cite {kurcsunoguglu1951space}

\item miscellaneous: \cite{muto1953some,rosquist1980global,caviglia1982equation,dolan1984solutions,roy1988cyl,dryuma1997,2015GReGr}

\end{itemize}
This compilation consists of solutions for null or timelike trajectories for either general or special trajectories. The solution methods vary greatly, ranging from direct integration to application of Newman-Penrose scalars. Nevertheless, derivation of BGOs for these spacetimes is still an open problem, which is alleviated by the existence of these solutions.

\chapter{Addendum: list of exact expressions of Synge's world function}

Synge's world function satisfies a couple of inocuously looking nonlinear partial differential equations (actually, it is the same equation which has to hold wrt coordinates at both endpoints) whose solution 
depends on the number of variables twice the dimension of the underlying space. Not surprisingly, symmetries simplify 
the problem. For example, in static spaces one can simply track the difference of coordinate times. In spherically 
symmetric spaces one can replace individual angular variables by a spherical angle. Nevertheless, this does not 
guarantee that the final expression will be tractable. It may happen that one is left with a system of transcendental 
equations involving nontrivial integrals. 

On the other hand, there have been very few attempts to study even the simpler cases or lower-dimensional toy models. 
Moreover, in the branch of classical optics there is a concept of the Hamilton's point characteristic function whose square is 
the world function, and there have been attempts to calculate it for some optical systems. Deeper inspection of 
this branch may yield additional insight into relativistic applications, but this is out of the scope of this 
thesis. Below we provide a list of spacetimes for which the Synge's world function has been studied analytically.

\begin{itemize}

\item Minkowski space: \cite{ruse1930}

\item Subcases of FLRW \cite{ruse1930, buchdahl1972, roberts1993} including Einstein static universe, generalized Milne universe, (anti-) de Sitter spacetime, FLRW with massless scalar field, Tolman radiation universe

\item  pp-wave: \cite{gunther1965, harte2012, harte2013}

\item Gödel universe: \cite{warner1980}
    
\item Schwarzschild spacetime (as a Taylor expansion): \cite{buchdahl1970, buchdahl1979, buchdahl1983}

\item other static spherically symmetric spacetimes \cite{buchdahl1970, john1984, john1989} including closed/open de Sitter, closed Einstein and Bertotti-Robertson universes


\end{itemize}

\end{document}